\pdfoutput=1
\documentclass{PoS}

\usepackage{xspace}
\usepackage{subfig}
\usepackage{placeins}
\usepackage{siunitx}
\usepackage{float}
\usepackage{cite}
\usepackage{txfonts}
\usepackage{xcolor}
\usepackage{lineno}

\newcommand{\mll}{\ensuremath{m_{e\mu}}\xspace}

\newcommand{\dphill}{\ensuremath{\Delta\phi_{e\mu}}\xspace}
\newcommand{\ptlzero}{\ensuremath{p_{\text{T}}^{\text{lead~}\ell}}\xspace}

\def\WW{\ensuremath{WW}\xspace}

\def\f0{\ensuremath{f_{\mathrm{0}}}\xspace}
\def\fLR{\ensuremath{f_{\mathrm{L}} - f_{\mathrm{R}}}\xspace}
\newcommand {\llnn} {$\ell\ell\nu\nu$}
\newcommand {\xsfid} {\ensuremath{\sigma_{ZZ\to \ell\ell\nu\nu}^{\mathrm{fid.}}}\xspace}
\newcommand {\ptll} {\ensuremath{p_{\mathrm{T}}^{\ell\ell}}\xspace}
\newcommand*{\mFourL}{\ensuremath{m_{4\ell}}\xspace}

\newcommand*{\WWW}{\ensuremath{WWW}\xspace}

\newcommand*{\WWZ}{\ensuremath{WWZ}\xspace}

\newcommand*{\WZZ}{\ensuremath{WZZ}\xspace}
\newcommand*{\WVZ}{\ensuremath{WVZ}\xspace}
\newcommand*{\WWg}{\ensuremath{WW \gamma}\xspace}
\newcommand*{\Wgg}{\ensuremath{W \gamma \gamma}\xspace}
\newcommand*{\WZg}{\ensuremath{W Z \gamma}\xspace}

\newcommand*{\wz}{\ensuremath{WZ}\xspace}

\newcommand{\elel}{\ensuremath{ee}\xspace}
\newcommand{\elmu}{\ensuremath{e\mu}\xspace}
\newcommand{\muel}{\ensuremath{\mu e}\xspace}
\newcommand{\mum}{\ensuremath{\mu\mu}\xspace}
\newcommand{\ssll}{\ensuremath{\ell \nu \ell \nu qq}\xspace}
\newcommand{\threelep}{\ensuremath{\ell \nu \ell \nu \ell \nu}\xspace}
\newcommand*{\TLjone}{3$\ell$-1j\xspace}
\newcommand*{\TLjtwo}{3$\ell$-2j\xspace}
\newcommand*{\TLjthree}{3$\ell$-3j\xspace}
\newcommand*{\FLDF}{4$\ell$-DF\xspace}
\newcommand*{\FLSFZ}{4$\ell$-SF-Z\xspace}
\newcommand*{\FLSFnoZ}{4$\ell$-SF-noZ\xspace}

\newcommand{\sswww}{\ensuremath{WWW \to \ell \nu \ell \nu qq}\xspace}

\newcommand{\lllwvz}{\ensuremath{WVZ \to \ell \nu qq \ell \ell}\xspace}
\newcommand{\llllwwz}{\ensuremath{WWZ \to \ell \nu \ell \nu \ell \ell}\xspace}
\newcommand{\llllwzz}{\ensuremath{WZZ \to qq \ell \ell \ell \ell}\xspace}

\newcommand{\lljj}{\ssll}

\newcommand*{\obsvvverr}{\ensuremath{1.38^{+0.39}_{-0.37}}\xspace}
\newcommand*{\ggZZ}{\ensuremath{gg\to  4\ell}\xspace}
\newcommand*{\qqZZ}{\ensuremath{q\bar{q}\to 4\ell}\xspace}

\newcommand {\shat} {\ensuremath{\sqrt{\hat{s}}}\xspace}
\newcommand*{\ZFourL}{\ensuremath{Z\to 4\ell}\xspace}
\newcommand{\ggB}{\ensuremath{gg\to 4\ell}}
\newcommand*{\ptFourL}{\ensuremath{\pt^{4\ell}}\xspace}
\newcommand {\VT} {$V_{\!\mathrm{T}}$}
\newcommand {\ST} {$S_{\!\!\mathrm{T}}$}
\newcommand {\VToST} {$V_{\!\mathrm{T}} / S_{\!\!\mathrm{T}}$}
\newcommand {\nonresll} {non-resonant-\ensuremath{\ell\ell}}

\newcommand*{\POWHEGBOX}{\textsc{Powheg-Box}\xspace}
\newcommand*{\POWHEG}{\textsc{Powheg}\xspace}
\newcommand*{\SHERPA}{\textsc{Sherpa}}
\newcommand*{\PYTHIA}{\textsc{Pythia}}
\def\powhegpythia{\textsc{Powheg+Pythia}\xspace}
\def\matrix{\textsc{MATRIX}\xspace}
\def\MATRIX{\textsc{MATRIX}\xspace}

\newcommand*{\antibar}[1]{\ensuremath{#1\bar{#1}}\xspace}

\newcommand*{\ttbar}{\antibar{t}}

\newcommand*{\ttZ}{\ensuremath{\ttbar Z}\xspace}

\newcommand*{\TeV}{\ifmmode {\mathrm{\ Te\kern -0.1em V}}\else
                   \textrm{Te\kern -0.1em V}\fi}%
\newcommand*{\GeV}{\ifmmode {\mathrm{\ Ge\kern -0.1em V}}\else
                   \textrm{Ge\kern -0.1em V}\fi}%
\newcommand*{\pt}{\ensuremath{p_{\rm T}}\xspace}
\newcommand*{\pT}{\ensuremath{p_{\rm T}}\xspace}

\newcommand{\stat}{\ensuremath{\:\textrm{(stat.)}}}
\newcommand{\syst}{\ensuremath{\:\textrm{(syst.)}}}
\newcommand{\lumi}{\ensuremath{\:\textrm{(lumi.)}}}

\newcommand*{\HT}{\ensuremath{H_{\mathrm{T}}}\xspace}
\newcommand*{\MET}{\ensuremath{E_{\mathrm{T}}^{\mathrm{miss}}}\xspace}
\newcommand*{\met}{\ensuremath{E_{\mathrm{T}}^{\mathrm{miss}}}\xspace}

\renewcommand*{\to}{\ensuremath{\rightarrow}\xspace}

\newcommand{\coll}[1]{#1 Collaboration}

\newcommand{\arxiv}[2]{arXiv:\href{http://www.arxiv.org/abs/#1}{\color[rgb]{0.,0.7,0.}{#1 [hep-#2]}}}
\newcommand{\pub}[2]{\href{http://dx.doi.org/#2}{\color[rgb]{0.,0.7,0.}{#1}}}
\newcommand{\subm}[1]{submitted to #1}
\newcommand{\accept}[1]{accepted for publication in #1}

\newcommand{\EPJC}{Eur.\ Phys.\ J.\ C}
\newcommand{\PLB}{Phys.\ Lett.\ B}

\title{Recent diboson and multiboson results in ATLAS}

\ShortTitle{Diboson and multiboson results} 

\author{\speaker{Markus Cristinziani}\thanks{Supported by the European Research
Council grant ERC--CoG--617185 and by the German Federal Ministry of Education and 
Research (FSP-103)}\\
{\rm On behalf of the ATLAS Collaboration}\\
Physikalisches Institut, Universit\"at Bonn, Nussallee 12, 53115 Bonn, Germany.\\
E-mail: \email{cristinz@uni-bonn.de}}

\abstract{%
Recent measurements of the associated production of two or three massive vector
bosons in proton--proton collisions at the Large Hadron Collider, collected by
the ATLAS detector at a centre-of-mass energy of $\!\sqrt{s} = 13 \TeV$, are
reported.  The diboson analyses target the production and decay modes
$WW\rightarrow e\nu_e\mu\nu_\mu$, $WZ \rightarrow \ell^{\prime} \nu \ell
\ell\,(\ell = e, \mu)$, $ZZ \rightarrow \ell\ell\nu\nu$, and the $4\ell$ final
state.  Fiducial inclusive cross sections are compared to state-of-the-art
Standard Model calculations and several unfolded differential distributions are
measured. In the $WZ$ channel, the helicities of the $W$ and $Z$ bosons are
probed.  Results are interpreted in terms of limits on anomalous gauge
couplings in the framework of effective field theories.  Finally, the first
evidence for the production of three massive vector bosons ($WWW$, $WWZ$,
$WZZ$) is presented.
}

\FullConference{7th Annual Conference on Large Hadron Collider Physics --- LHCP2019\\
		20--25 May, 2019\\
		Puebla, Mexico}

\begin{document}

\section{Introduction}

Five recent measurements of the ATLAS
Collaboration~\cite{WW,WZ,ZZinv,FourLep,VVV} are presented of the production of
two or three massive vector bosons $V$ ($V = W$ or $Z$) in proton--proton
($pp$) collisions at a centre-of-mass energy of $\sqrt{s} = 13 \TeV$ at the
Large Hadron Collider (LHC).  The diboson production measurements of
$WW$~\cite{WW}, $WZ$~\cite{WZ}, $ZZ$~\cite{ZZinv}, and $ZZ$ along with $Z\to4\ell$
and $H\to 4\ell$~\cite{FourLep} have been performed using the data collected in 2015 and
2016, corresponding to a dataset of \SI{36.1}{\per\femto\barn}, while the
measurement targeting the production of three bosons, $WWW$, $WWZ$,
$WZZ$~\cite{VVV}, are based on a larger dataset of \SI{79.8}{\per\femto\barn},
i.e.  including 2017 data.

The study of the production of multiple massive vector bosons through
interactions of quarks and gluons probes the electroweak (EW) non-Abelian gauge
structure of the Standard Model (SM) and allows further tests of the strong
interaction between quarks and gluons. It provides the means to directly probe
triple (TGC) and quartic (QGC) gauge boson couplings.  On the other hand, a
broad range of new phenomena beyond the SM (BSM) are predicted to reveal
themselves through multiboson production.  Improved constraints from precise
measurements can potentially probe scales of new physics in the multi-\TeV\
range and provide a way to look for signals of new physics in a
model-independent way.  

Sizeable production cross sections combined with the large sample of $pp$
collision data delivered by the LHC, enable these processes to be studied with
a better statistical precision than was possible in previous measurements or
establish them for the first time.  While $W$ bosons are produced more
copiously than $Z$ bosons, the signal-to-background ratio is in general more
advantageous in channels containing leptonically-decaying $Z$ bosons and
therefore the final states and analyses have similar sensitivities to uncover
departures from the SM predictions.

\section{Measurements of two massive vector bosons using 2015 and 2016 data}
\subsection{Production of $WW\rightarrow e\nu_e\mu\nu_\mu$}

The cross section for the production of $W^+W^-$ pairs is measured in a
fiducial phase space selecting one electron and one muon, targeting the decay
$WW\rightarrow e^{\pm}\nu\mu^{\mp}\nu$. The fiducial phase space is chosen to
be orthogonal to the ATLAS $H \to WW$ measurements by means of a requirement on
the dilepton invariant mass.  Events with a same-flavour lepton pair or with
jets with a transverse momentum (\pt) above $35 \GeV$ are discarded because they have
a larger background from the Drell--Yan process, or from top quarks,
respectively. Additional requirements are imposed in order to reduce the
dominant backgrounds.  After event selection, the signal-to-background ratio is
approximately $2$, with top-quark being the largest background, followed by
Drell--Yan, $W$+jets and $WZ$ events.  The systematic uncertainty for the
inclusive cross-section measurement amounts to $7\,\%$, dominated by the
uncertainties related to the identification of $b$-quark-initiated jets ($b$-tag), the
estimation of the $W$+jets background, and the uncertainties related to the
identification of jets.

\begin{figure}[htbp]
\centering
\hfill
\subfloat[]{\raisebox{1em}{\includegraphics[height=0.27\textheight]{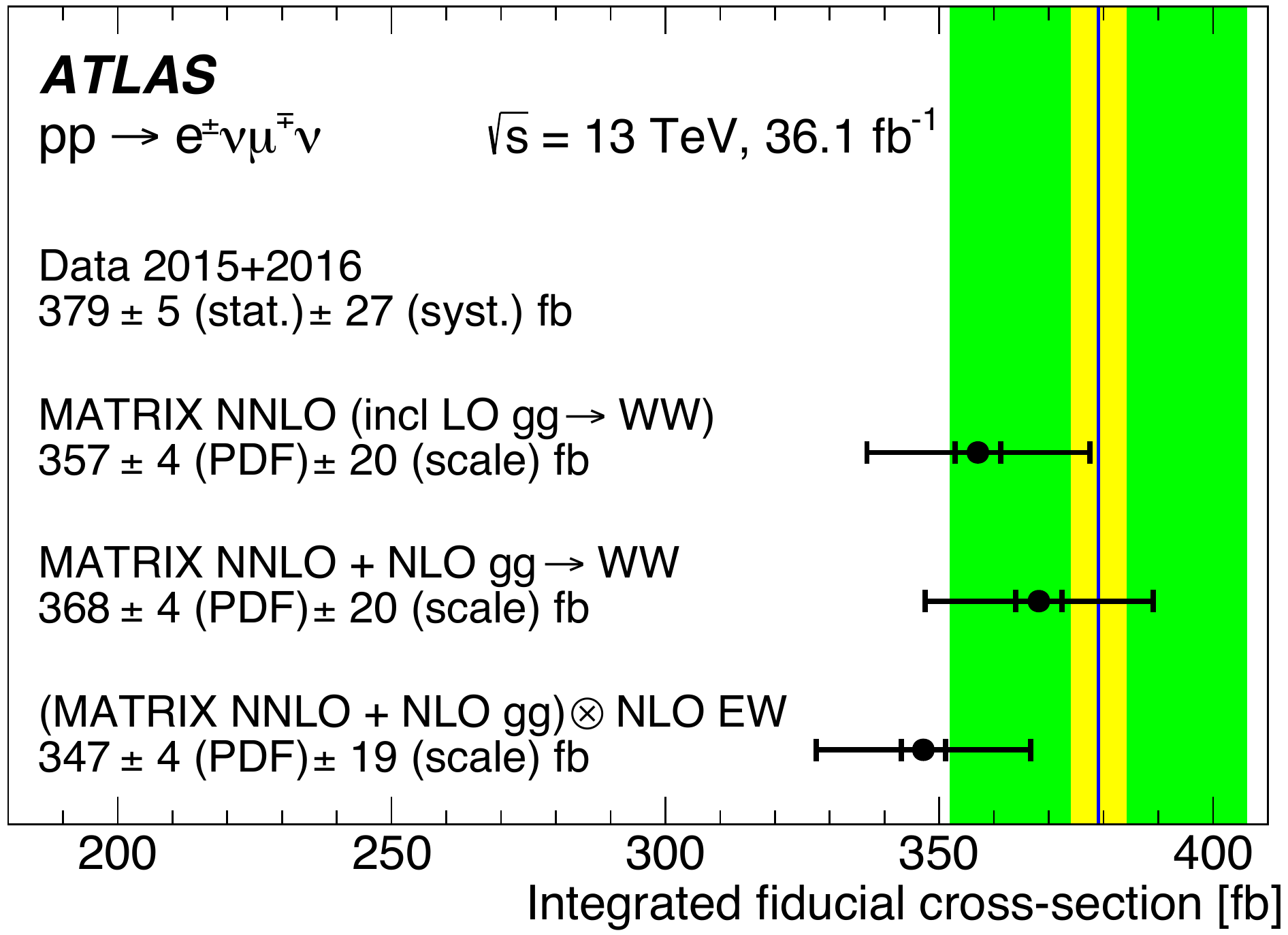}}}
\hfill \hfill
\subfloat[]{\includegraphics[height=0.30\textheight]{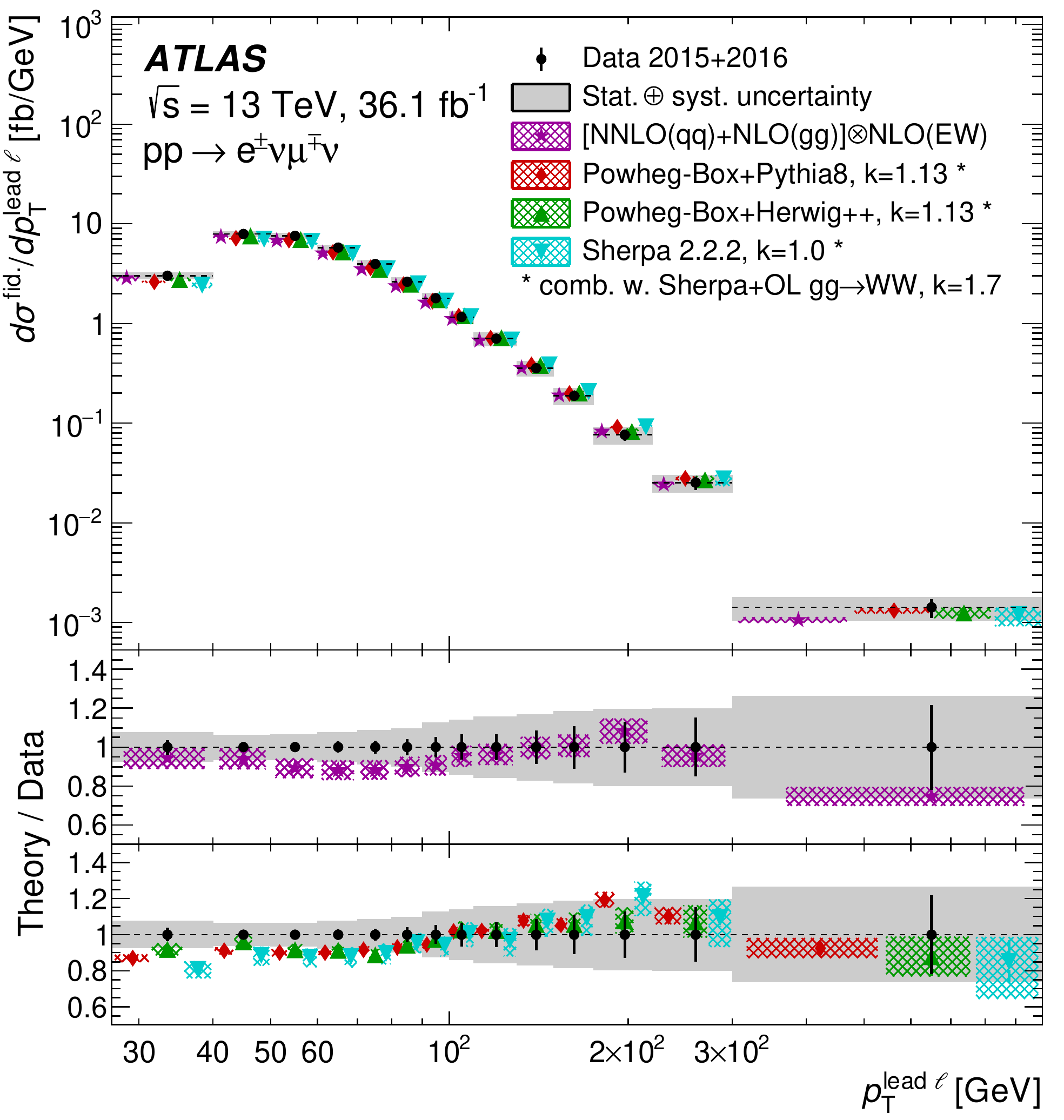}}
\hfill
\caption{(a) Comparison of the measured fiducial cross section, indicated
by a vertical line and bands, with various theoretical predictions, indicated
as points with error bars~\cite{WW}.  (b) Measured fiducial cross section as a
function of \ptlzero, compared to corrected NNLO predictions and various
Monte-Carlo simulations~\cite{WW}.  Theoretical uncertainties correspond to PDF+scale
variations.}
\label{fig:WW}
\end{figure}

The measured fiducial cross section is found to be consistent with theoretical
predictions, including next-to-next-to leading order (NNLO) QCD and next-to
leading order (NLO) EW corrections (Figure~\ref{fig:WW}a).  The result is:
$\sigma^\text{fid.}_{WW\rightarrow e\nu_e\mu\nu_\mu}= 379.1 \pm 5.0 \stat \pm
25.4 \syst \pm 8.0 \lumi$ fb.  The fiducial cross section is also measured as a
function of the \pt threshold for the jet veto, where the
fiducial cross section rises by about $30\,\%$ when accepting events containing
jets with a \pt of up to $60 \GeV$, as compared with $30 \GeV$.
Predictions agree within uncertainties with the data, but are consistently at
the lower bound of these and a slight slope in the ratio of predictions to data
is observed.

Six differential distributions involving kinematic variables of the final-state
leptons are measured. Three of them characterise the energy of the process: the
transverse momentum of the leading lepton \ptlzero, the invariant mass \mll\
and the transverse momentum of the dilepton system. Three further distributions
probe angular correlations and the spin state of the \WW\ system: the rapidity
of the dilepton system, the difference in azimuthal angle between the decay
leptons \dphill, and $|\tanh((\Delta \eta_{e\mu})/2)|$, where $\Delta
\eta_{e\mu}$ is the difference between the pseudorapidities of the leptons.
The differential cross sections are compared with several predictions from
perturbative QCD calculations.  Data and theory show fair agreement for all
distributions, except for low values of $m_{e\mu}$, for $\dphill < 1.8$, as
well as for low values of \ptlzero (Figure~\ref{fig:WW}b).

\FloatBarrier

\subsection{Production of $WZ \rightarrow \ell^{\prime} \nu \ell \ell$}

The $WZ$ production cross section is measured within a fiducial phase space
closely matching the detector acceptance, both inclusively and differentially
as a function of several individual variables related to the kinematics of the
\wz\ system and to the jet activity in the event.  The $W$ and $Z$ bosons are
reconstructed using their decay modes into electrons or muons, employing the
so-called resonant-shape algorithm. All final states with three leptons and
missing transverse momentum (\MET) are considered, and categorised into the
channels $eee$, $\mu e e$, $e \mu\mu$ and $\mu\mu\mu$, where the first lepton
is the one assigned to the $W$ boson decay.  No requirement on the number of
jets is applied.  After event selection, the signal-to-background ratio is
approximately $4$ and the dominant backgrounds are processes with misidentified
leptons and $ZZ$ production.  Systematic uncertainties are considerably reduced
by combining the different final states.  The dominant uncertainty is related
to the estimation of the contribution by the misidentified leptons.

\begin{figure}[htbp]
\centering
\hfill
\subfloat[]{\raisebox{1em}{\includegraphics[height=0.28\textheight]{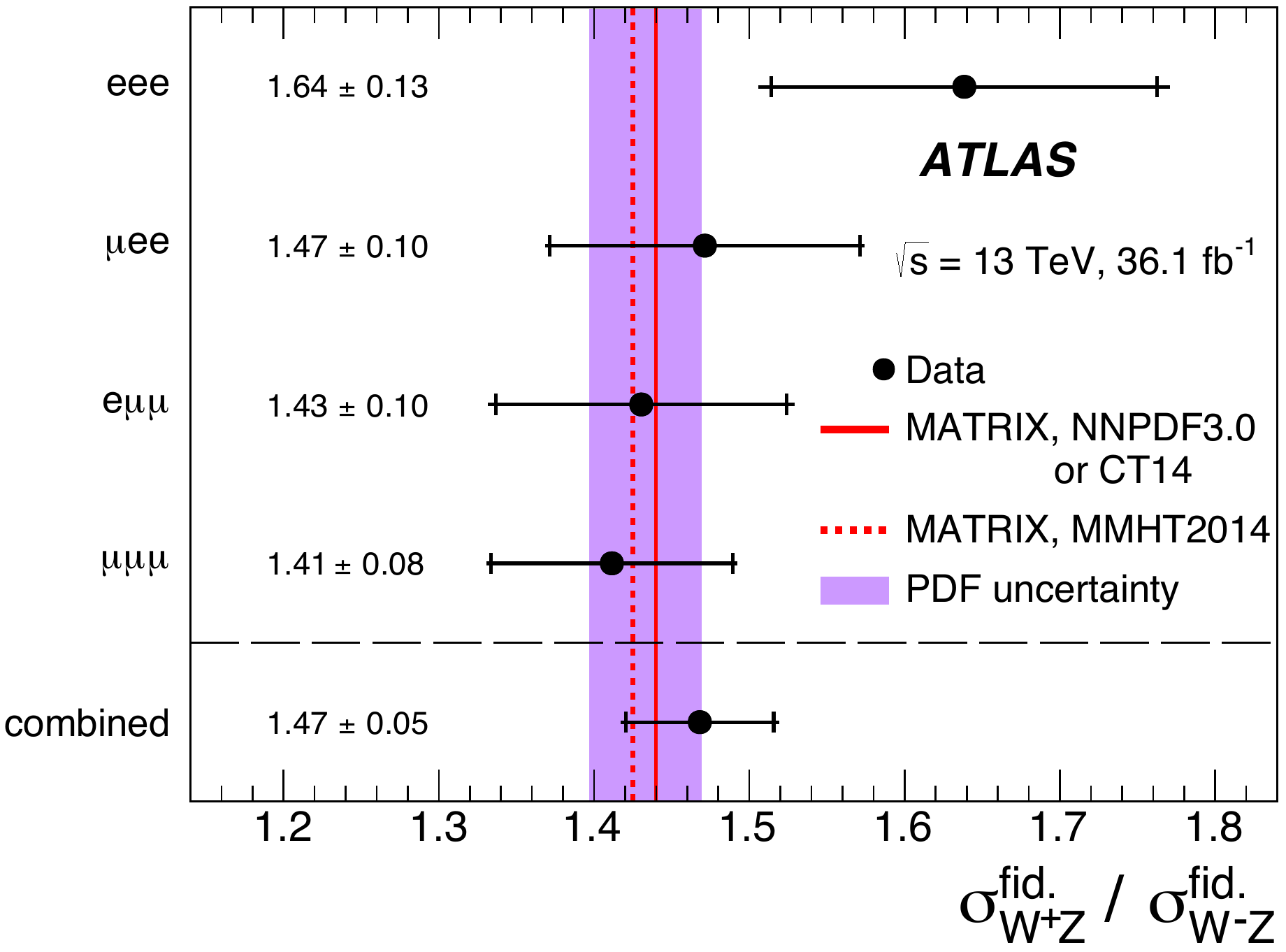}}}
\hfill\hfill
\subfloat[]{\includegraphics[height=0.31\textheight]{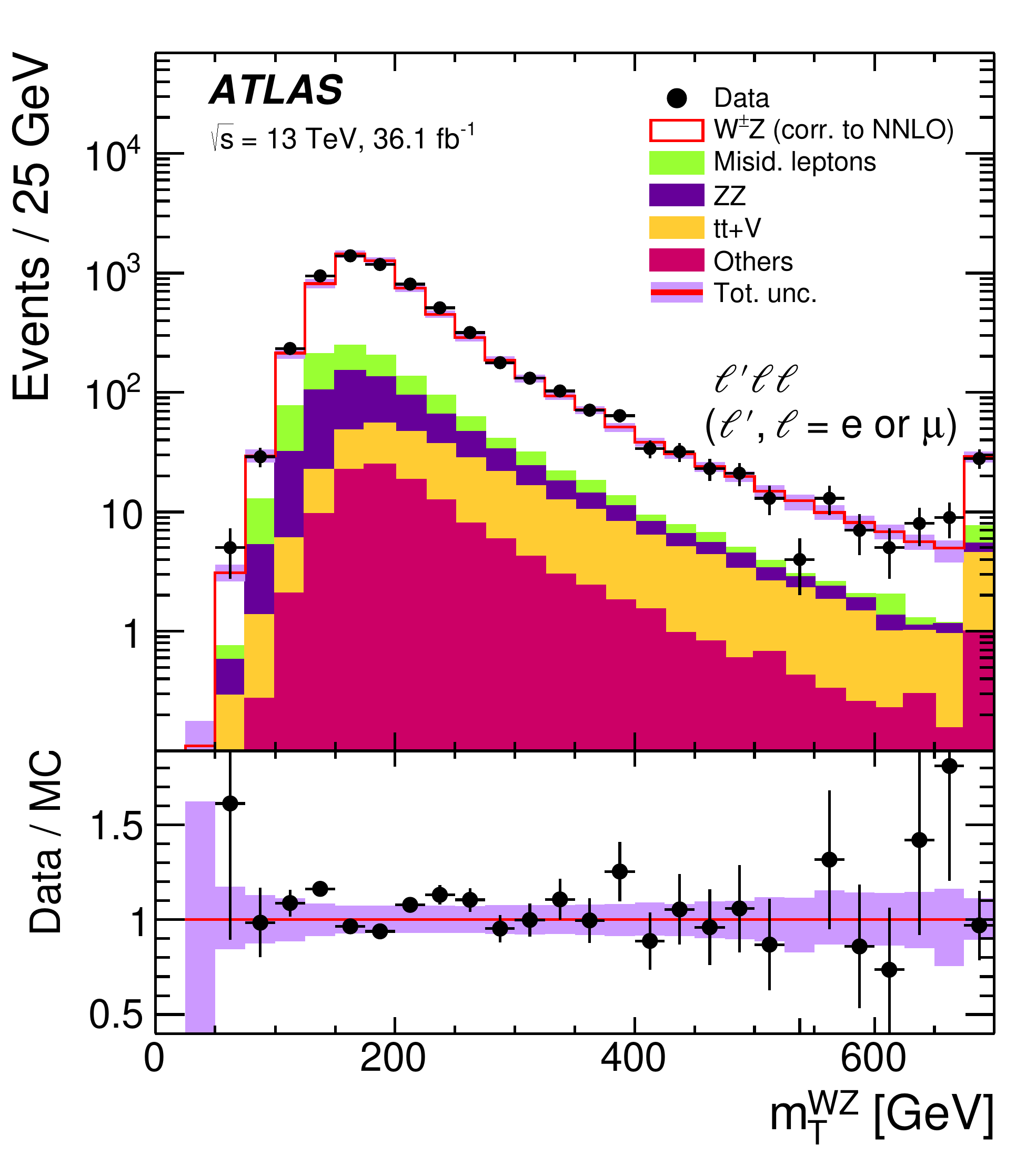}}
\hfill
\caption{%
(a) Measured ratio $\sigma^{\mathrm{fid.}}_{W^{+}Z} /
\sigma^{\mathrm{fid.}}_{W^{-}Z}$ of $W^{+}Z$ and $W^{-}Z$ integrated cross
sections in the fiducial phase space in each of the four channels and for the
combination~\cite{WZ}. (b) Distribution of the kinematic variable $m_\textrm{T}^{WZ}$~\cite{WZ}.}
\label{fig:WZone}
\end{figure}

The measured inclusive cross section in the fiducial region is $\sigma_{W^\pm Z
\rightarrow \ell^{\prime} \nu \ell \ell}^{\textrm{fid.}} = 63.7 \pm 1.0 \stat
\pm 2.3 \syst \pm 1.4 \lumi$ fb, in agreement with the NNLO Standard Model
expectation of $61.5^{+1.4}_{-1.3}$~fb.  The ratio of the $W^+Z$ cross section
to the $W^-Z$ cross section, which is sensitive to the parton distribution
functions (PDF), is also measured. Here, most of the systematic uncertainties
almost cancel out in the ratio.  The result, $\sigma^{\mathrm{fid.}}_{W^{+}Z} /
\sigma^{\mathrm{fid.}}_{W^{-}Z} = 1.47 \pm 0.05 \stat \pm 0.02 \syst$, is
compared with the SM calculation and several PDF sets in
Figure~\ref{fig:WZone}a.

The $WZ$ production cross section is measured as a function of several
kinematic variables and compared with SM predictions at NNLO from the \matrix
calculation and at NLO from the \powhegpythia and \SHERPA\ event generators:
the transverse momentum of the $Z$ boson, the transverse momentum of the $W$
boson, the transverse mass of the $WZ$ system $m_\textrm{T}^{WZ}$
(Figure~\ref{fig:WZone}b), the azimuthal angle between the $W$ and $Z$ bosons
$\Delta \phi (W,Z)$, the \pt of the neutrino associated with the decay of the
$W$ boson, and, finally, the absolute difference between the rapidities of the
$Z$ boson and the lepton from the decay of the $W$ boson.  The differential
cross-section distributions are well described by the theory predictions, with
the exception of the jet multiplicity.  The \MATRIX calculations show the best
agreement with the data, for instance for the $\Delta \phi (W,Z)$ distribution
which is sensitive to QCD higher-order perturbative effects.

An analysis of angular distributions of the leptons from the decays of the $W$
and $Z$ bosons has been performed. The normalised differential distribution as
a function of $\cos \theta_{\ell, V}$ depends on the longitudinal ($f_0$),
transverse left-handed ($f_L$) and transverse right-handed ($f_R$) helicity
fractions, where $\theta_{\ell, V}$ is defined using the helicity frame, as the
decay angle of the lepton in the $V$ boson rest frame relative to the $V$
direction in the $WZ$ centre-of-mass frame.  The existence of the
longitudinally polarised state is a consequence of the non-vanishing mass of
the bosons generated by the electroweak symmetry breaking (EWSB) mechanism.
The measurement of the polarisation in diboson production therefore tests both
the SM innermost gauge symmetry structure, through the existence of TGC, and
the particular way this symmetry is spontaneously broken, via the longitudinal
helicity state.  Angular observables can be used to look for new interactions
that can lead to different polarisation behaviour than predicted by the SM, to
which the \wz final state would be particularly sensitive.  

Helicity fractions  of pair-produced vector bosons are measured for the first
time in hadronic collisions.  The three helicity fractions of the $W$ and $Z$
bosons are measured using a template fit to the $\cos \theta_{\ell, V}$
distributions.  Integrated over the fiducial region, the longitudinal
polarisation fractions of the $W$ and $Z$ bosons in $WZ$ events are measured to
be $f_0^W = 0.26 \pm 0.06$ and $f_0^Z = 0.24 \pm 0.04$, in agreement with the
SM predictions at NLO in QCD and at leading order (LO) for EW corrections, of
$0.238 \pm 0.003$ and $0.230 \pm 0.003$, respectively. The observed
significances of the measurements are $4.2\,\sigma$ and $6.5\,\sigma$ for
$f_0^W$ and $f_0^Z$, respectively.  The measurements are dominated by
statistical uncertainties.  

\begin{figure}[htbp]
\centering
\hfill
\subfloat[]{\includegraphics[height=0.28\textheight]{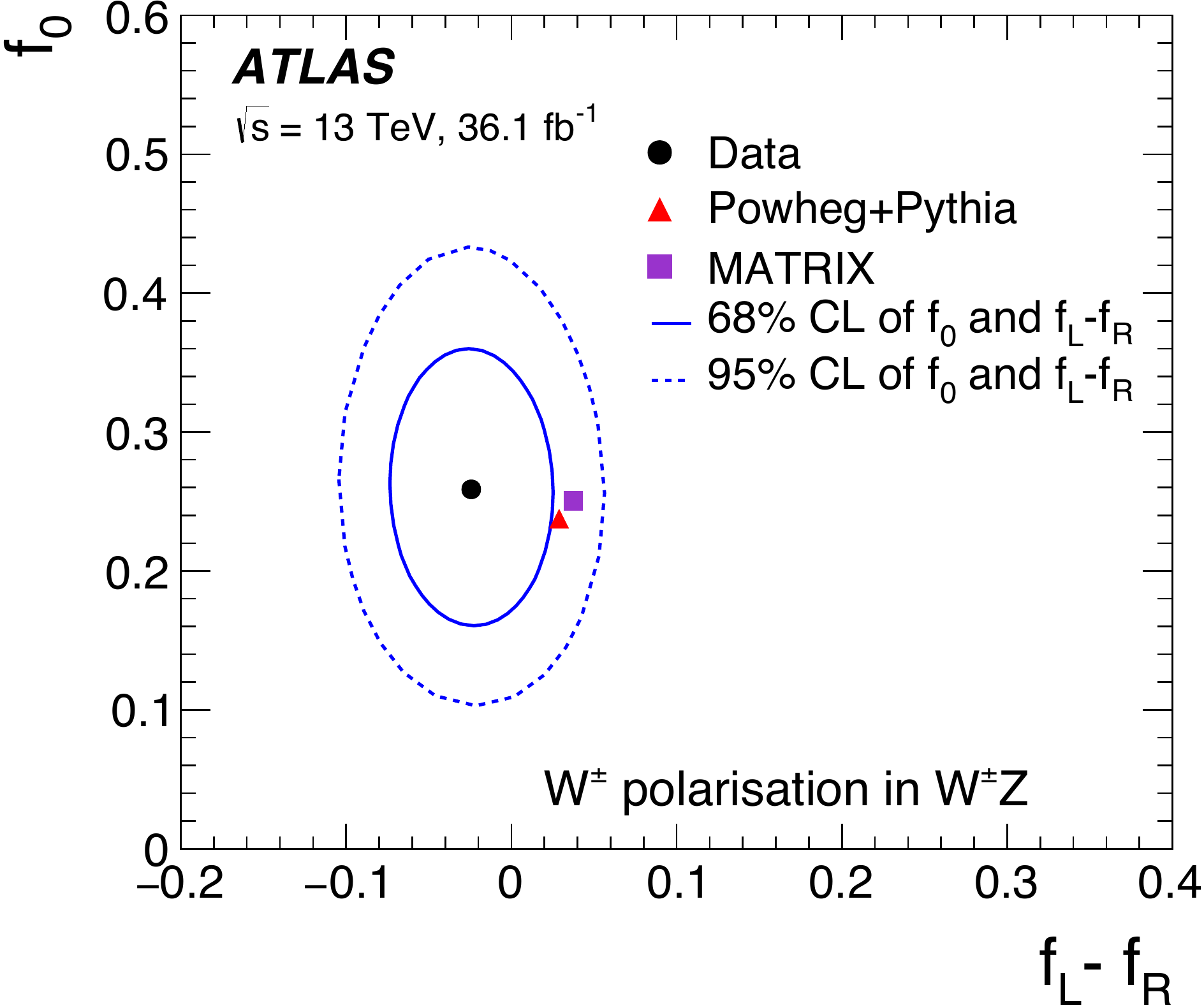}}
\hfill\hfill
\subfloat[]{\includegraphics[height=0.28\textheight]{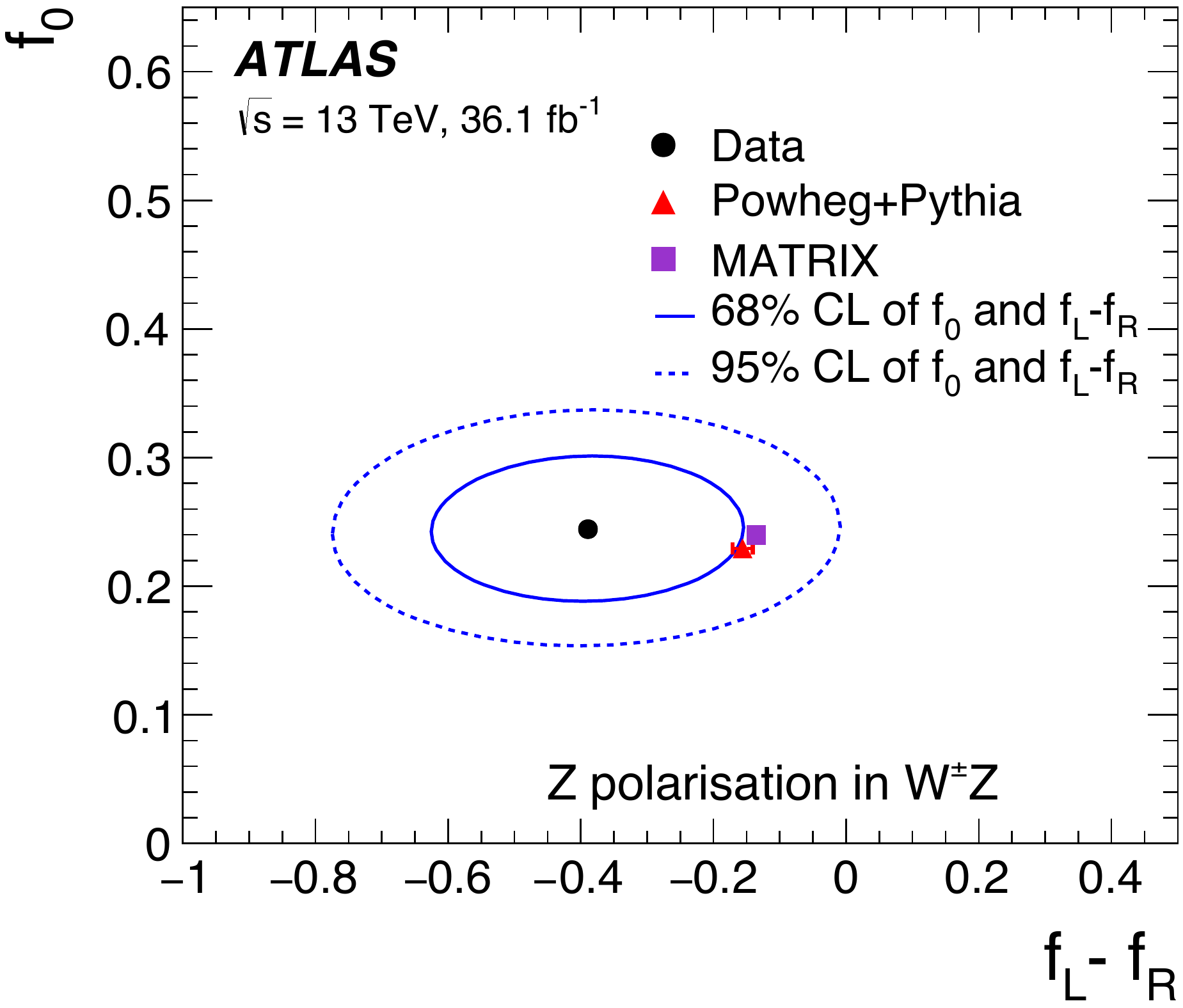}}
\hfill
\caption{
Measured helicity fractions \f0 and \fLR for the (a) $W$ boson and (b) $Z$ boson in
$W^{\pm}Z$ events, compared with predictions at LO for the electroweak
interaction and with $\sin^2{\theta_{\mathrm{W}}}=0.23152$ from \powhegpythia
and \matrix~\cite{WZ}.}  
\label{fig:WZtwo}
\end{figure}

The differences of the left and right transverse polarisations are also
measured, with the constraint $\f0+f_L+f_R=1$.  The values of \f0 and \fLR,
measured in \wz events, are shown in Figure~\ref{fig:WZtwo}.  They agree with
the predictions from \powhegpythia and \matrix within less than one and two
standard deviations of their uncertainties for \f0 and \fLR, respectively. No
stringent constraints nor clear inconsistencies between measurements and
predictions can be deduced.  Polarisation measurements for each charge of the
$W$ boson might be helpful in the investigation of $CP$ violation effects in
the interaction between gauge bosons.  In the longer term, measuring the
scattering of longitudinally polarised vector bosons will be a fundamental test
of EWSB.

\FloatBarrier

\subsection{Production of $ZZ \rightarrow \ell\ell\nu\nu$} 

\begin{figure}[htbp]
\centering
\hfill
\subfloat[]{\includegraphics[height=0.25\textheight]{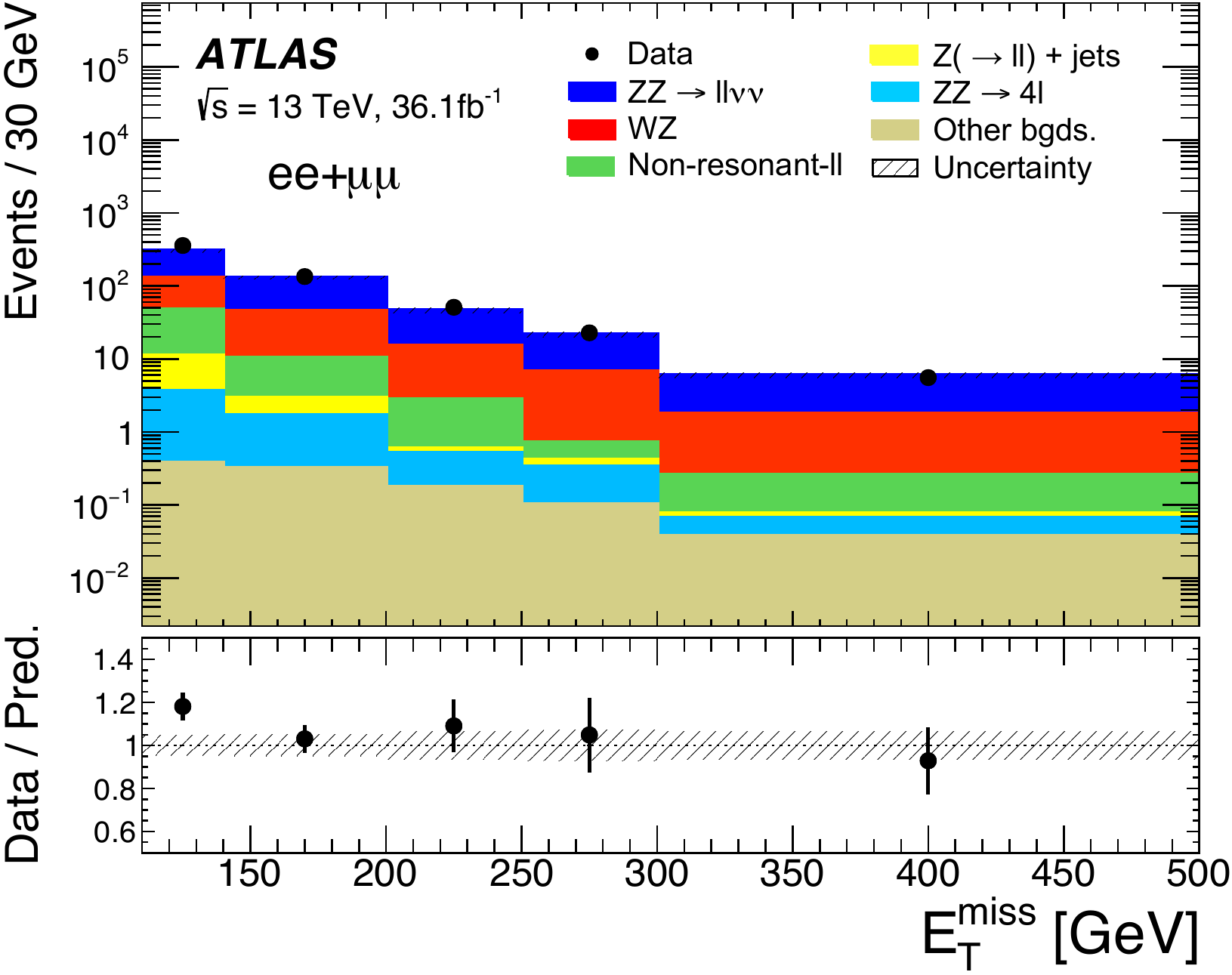}}
\hfill\hfill
\subfloat[]{\includegraphics[height=0.25\textheight]{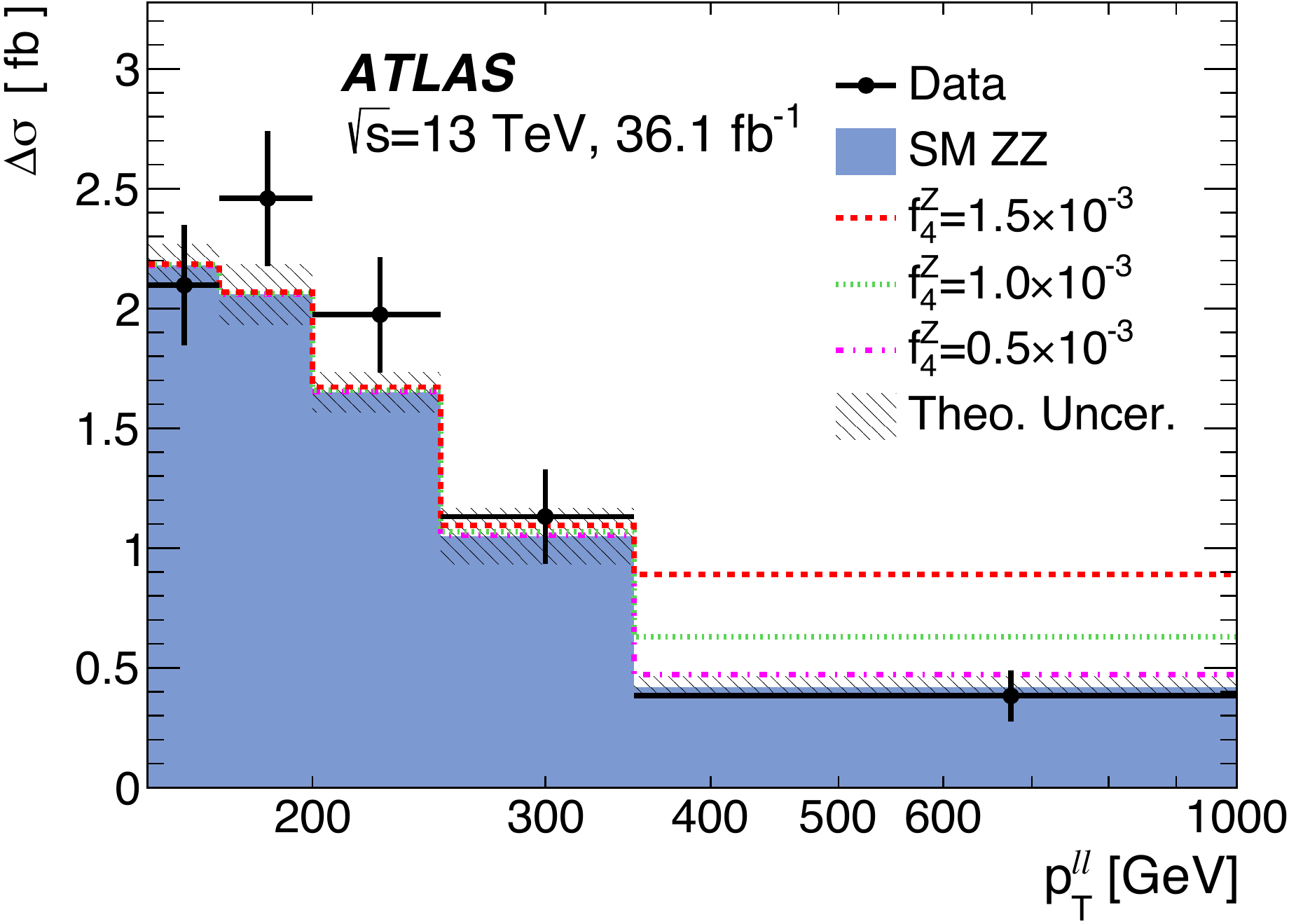}}
\hfill
\caption{(a) Observed and expected \MET distribution for the $ZZ$ analysis~\cite{ZZinv}.
(b) Unfolded measured distribution of \ptll, compared with the SM prediction
from \POWHEG to which are added possible aTGC contributions for different
values of the strength of the coupling parameters (here for $f_4^{Z}$)~\cite{ZZinv}.} 
\label{fig:ZZ}
\end{figure}

The $ZZ$ production in the \llnn\ final state has a larger branching fraction
but suffers from higher background contamination in comparison with the $ZZ \to
4\ell$ channel.  To ensure a good signal-to-background ratio, the experimental
selection requires one $Z$ boson boosted against the other in the transverse
plane, which results in a pair of high-\pt leptons and significant \met.  The
\llnn\ channel thus offers higher data statistics than the $4\ell$ channel for
events with high-\pt $Z$ bosons, and competitive precision for integrated and
differential measurements, as well as good sensitivity to anomalous TGC (aTGC).
The integrated cross section of $ZZ$ production is measured in a fiducial phase
space and then extrapolated to a total phase space.  The signal yield is
determined through a fit to the observed \met spectrum (Figure~\ref{fig:ZZ}a),
which leads to improved sensitivity compared with a simple event-counting
method.

Candidate events are selected by requiring exactly two electrons or muons with
opposite charges.  The dilepton invariant mass is required to be compatible
with the $Z$ boson mass.  Candidate events are required to have significant
\met and \VToST~$>0.65$, where \VT{} is the magnitude of the vector sum of
transverse momenta of selected leptons and jets, and \ST{} is the scalar \pT{}
sum of the corresponding objects. Additional selection criteria based on
angular variables are imposed to ensure the desired detector signature.
Finally, events containing one or more $b$-tagged jets are vetoed to further suppress
the \ttbar~and $Wt$ backgrounds.  After the event selection, the overall
signal-to-background ratio is about $1.7$.  The $WZ$ and \nonresll{} backgrounds
account for 72\% and 21\% of the total background contribution, respectively,
and are estimated from control regions in data. Systematic uncertainties are
dominated by the uncertainties in the background modelling followed by the
uncertainties in the jet identification.

The fiducial cross section of the combined $ee$ and $\mu\mu$ channels is
measured to be $\xsfid = 25.4 \pm 1.4 \stat \pm 0.9 \syst \pm 0.5 \lumi$ fb, in
agreement with the SM prediction of $22.4 \pm 1.3$ fb.  The integrated cross
sections in the total and fiducial phase spaces are measured with an
uncertainty of $7\%$, which is significantly better than the previous
measurement using the $8 \TeV$ data.  The measured cross sections is about
$13\%$ higher than the NNLO (quark-initiated) and NLO (gluon-initiated) SM
predictions for $ZZ$ production, but the difference is not significant
considering the measurement and prediction uncertainties.  

Differential cross sections are reported in the fiducial region for eight
kinematic variables, which are sensitive to effects from higher-order
corrections and possible BSM physics.  These variables include the transverse
momenta of the leading lepton, of the leading jet, of the dilepton system
(\ptll, Figure~\ref{fig:ZZ}b), and of the $ZZ$ system; the transverse mass of
the $ZZ$ system, the absolute rapidity of the dilepton system, the azimuthal
angle difference between the two leptons, and the number of jets.  Differential
cross sections are reported for these eight kinematic variables in the fiducial
phase space, and no significant deviation from the expectations is found.

\FloatBarrier

\subsection{Four-lepton production}

\begin{figure}[htbp]
\centering
\hfil
\subfloat[]{\raisebox{2.0em}{\includegraphics[height=0.24\textheight]{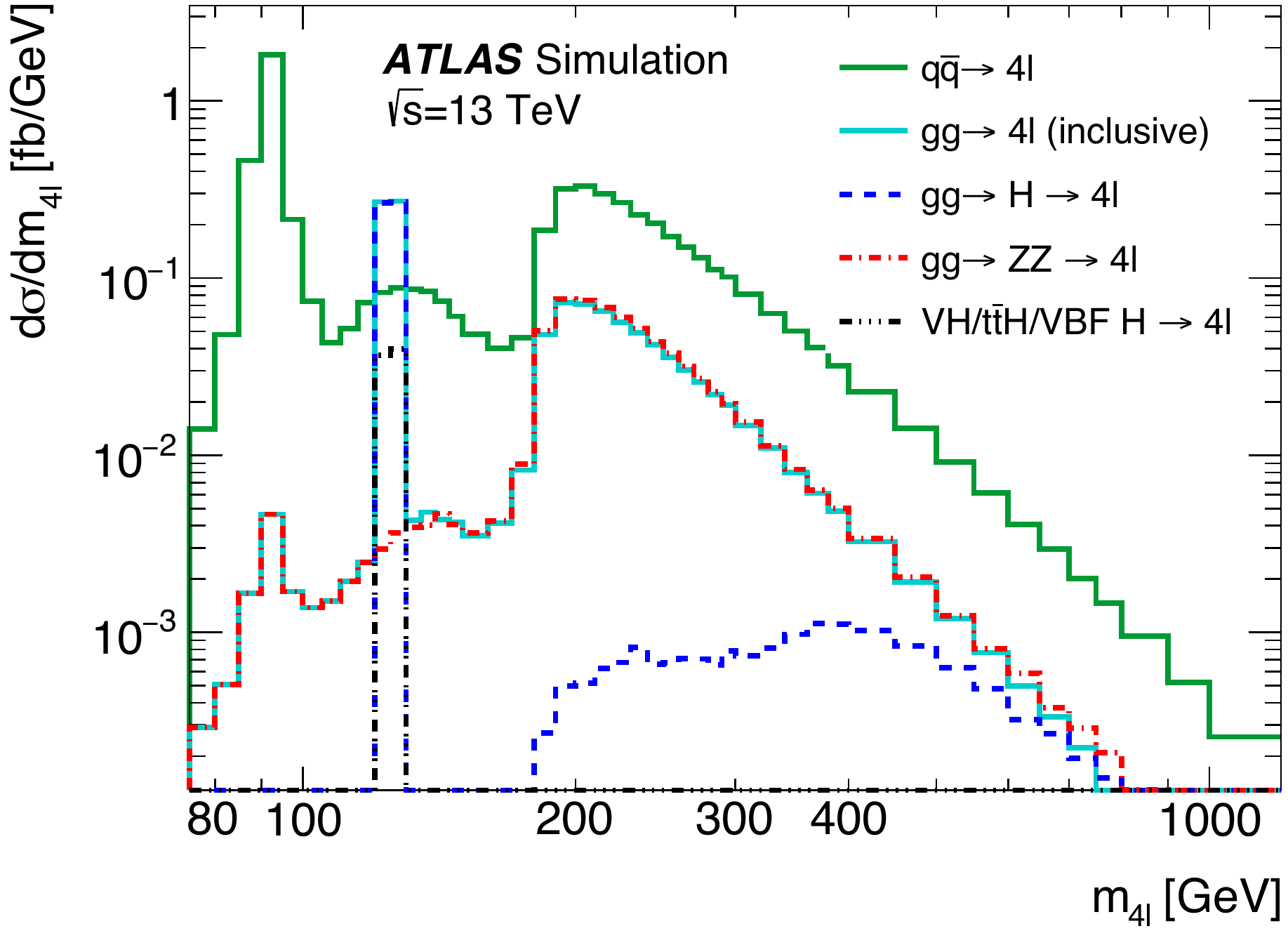}}}
\hfill \hfill
\subfloat[]{\includegraphics[height=0.30\textheight]{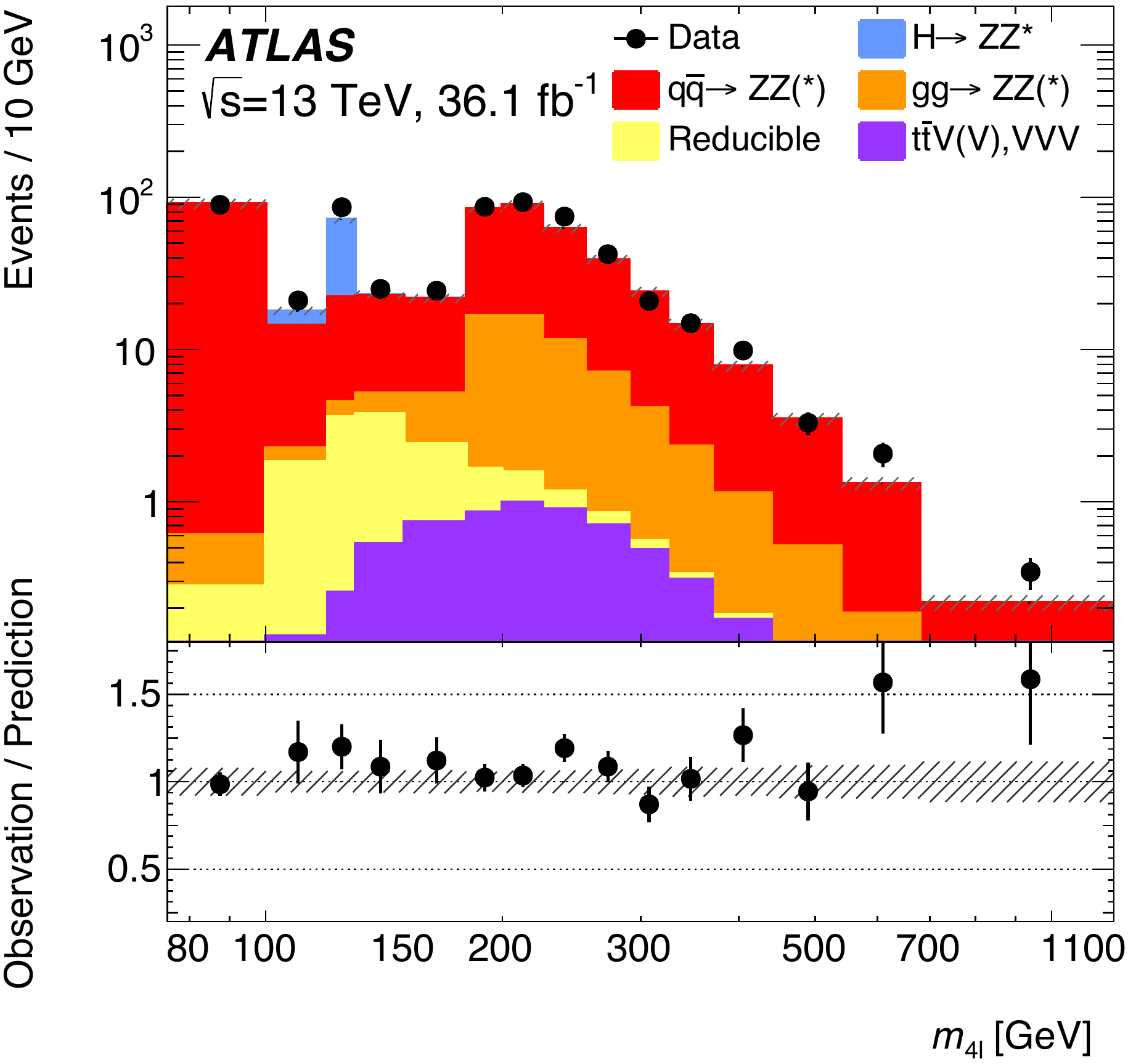}}
\hfill
\caption{(a) Differential cross sections as a function of the four-lepton
invariant mass $\mFourL$ predicted by simulation~\cite{FourLep}. The total $gg\rightarrow
4\ell$ includes contributions from $gg\rightarrow H^{(\ast)} \rightarrow 4\ell$
as well as $gg \rightarrow 4\ell$ and the interference between the two. The
$\qqZZ$ and $\ggZZ$ processes include off-shell Higgs boson production.
(b) Selected events as a function of $\mFourL$ compared to
the total SM prediction~\cite{FourLep}.}
\label{fig:FL}
\end{figure}

In $pp$ collisions, four-lepton production is expected to receive contributions
from several SM physics processes, resulting in a rich predicted cross-section
spectrum, shown as a function of the invariant four-lepton mass
$\mFourL$ in Figure~\ref{fig:FL}.  

Largest in magnitude is the quark-induced $t$-channel process \qqZZ, with
leptonic decays of the $Z$ bosons.  Gluon-induced \ggZZ production also occurs,
via an intermediate quark loop. The theoretical uncertainties in the SM
prediction for this latter contribution are comparatively large.  At around
$\mFourL\simeq m_Z$, single resonant \ZFourL production through QED radiative
processes leads to a peak in the spectrum, and allows an extraction of the
cross section and branching fraction for $Z \rightarrow 4\ell$ to be made.
Pairs of $Z$ bosons can also be produced from the decay of an intermediate
Higgs boson. The majority of these are produced via gluon--gluon fusion, with
minor contributions from vector-boson fusion and associated production with
vector bosons or top-quark pairs.  There is resonant production around the
Higgs boson mass, as well as off-shell production at higher mass values, which
is enhanced at approximately $350\GeV$ due to top-quark loops in the
gluon--gluon fusion mechanism.  At around $180\GeV$ there is an enhancement of
all the processes involving two $Z$ bosons, as on-shell production is possible
above this mass.  The box diagram processes $gg\rightarrow 4\ell$ and
$gg\rightarrow H^{(*)}\rightarrow 4\ell$ interfere destructively in the SM.
While interference is maximal around $\mFourL = 220\GeV$, the relative effect
of the $gg\rightarrow H^{(*)}\rightarrow 4\ell$ contribution to the overall
$\ggB$ lineshape is most pronounced above $350\GeV$, as is visible in
Figure~\ref{fig:FL}.  The off-shell Higgs production rate may be affected by
BSM processes involving additional heavy particles, or modifications of the
Higgs couplings, even if there is no effect on on-shell Higgs boson production.

Events are required to contain two pairs of same-flavour opposite-sign (SFOS)
leptons.  Contributions from leptonically decaying $\tau$-leptons and quarkonia
are reduced through requirements on the invariant masses of the dilepton pairs,
but otherwise the selection is rather inclusive.  The measurement is made
differentially in the invariant mass $\mFourL$ of the four-lepton system, and
double-differentially as a function of $\mFourL$ versus the transverse momentum
of the four-lepton system $\ptFourL$, the rapidity of the system, the
matrix-element discriminant $D_\textrm{ME}$ designed to isolate off-shell Higgs
boson contributions, and the final state lepton flavour channel.  The $\mFourL$
measurement is also made separately for each flavour combination of leptons in
the event: $4e$, $4\mu$ and $2e2\mu$.  The double-differential cross sections
can provide additional sensitivity to the various subprocesses contributing to
the measured final state. For example, the $\ptFourL$ is expected to
discriminate $gg\rightarrow ZZ$ from $q\bar{q}\to ZZ$. They are also of
interest for future interpretations, as some BSM contributions can have an
impact which depends upon the final-state lepton flavours.

The limiting source of uncertainty in this measurement is the statistical
uncertainty, which is many times larger than the total systematic uncertainty
in some bins. Experimental and theoretical sources both contribute to the
systematic uncertainty, and their relative impact varies depending on the bin.
The main systematic uncertainties are due to the lepton identification and
uncertainties related to the luminosity and pile-up.  The measurements are
consistent with the predictions of the SM.  The signal strength of the
gluon--gluon fusion production process is measured to be $\sigma_{\ggB}
/\sigma^{\text{SM}}_{\ggB} = 1.3 \pm 0.5$ compared to an expected value of $1.0
\pm 0.4$.  A value for the $Z \rightarrow 4\ell$ branching fraction of $(4.70
\pm 0.32 \stat \pm 0.25 \syst)\times 10^{-6}$ is obtained, consistent with
existing measurements and exceeding the precision of previous ATLAS results.
Finally, an upper limit on the signal strength for the off-shell Higgs
production process of $6.5$ is obtained.

\FloatBarrier

\subsection{Interpretation in terms of effective field theories}

New physics processes at a high energy scale $\Lambda$ that alter diboson
production can be described by operators with mass dimensions larger than four
in an effective field theory (EFT) framework.  The dimensionless coefficients
($c_i$) of the operators $\mathcal{O}_i$ parameterise the strength of the
coupling between new physics and SM particles, $\mathcal{L} =
\mathcal{L}_{\textrm{SM}}+ \sum_i^{}  \frac{c_i}{\Lambda^2}\mathcal{O}_i~$.
The higher-dimensional operators of the lowest order from purely EW processes
have dimension six, and can generate aTGCs.  A deviation from the SM in
measured production rates or in certain kinematic distributions, as predicted
by these theories, could provide evidence for BSM physics.

For the $WW$ measurement, in the EFT framework employed, there are five
dimension-six operators and the relevant EFT coefficients (coupling constants)
are: $c_{WWW}$, $c_{W}$, $c_{B}$, $c_{\tilde{W}WW}$ and $c_{\tilde{W}}$.  The
distribution of the transverse momentum of the leading lepton is used to
investigate aTGC parameters.  No evidence for anomalous $WWZ$ and $WW\gamma$
couplings is found, hence limits on their magnitudes are set.  Constraints on
the EFT coefficients are determined by considering one operator at a time using
the unfolded \ptlzero fiducial cross section.  Due to the higher centre-of-mass
energy, the limits observed are more restrictive than those previously
published by the ATLAS and CMS collaborations in the \WW\ final state.  The
sensitivity to dimension-six operators mostly stems from their direct effect on
the \WW\ cross section as a function of \ptlzero, except for the $c_{W}$
coefficient where both the direct contribution and the interference between the
SM and terms containing EFT operators contribute equally.

\begin{figure}[htbp]
\centering
\hfill
\subfloat[]{\includegraphics[height=0.25\textheight]{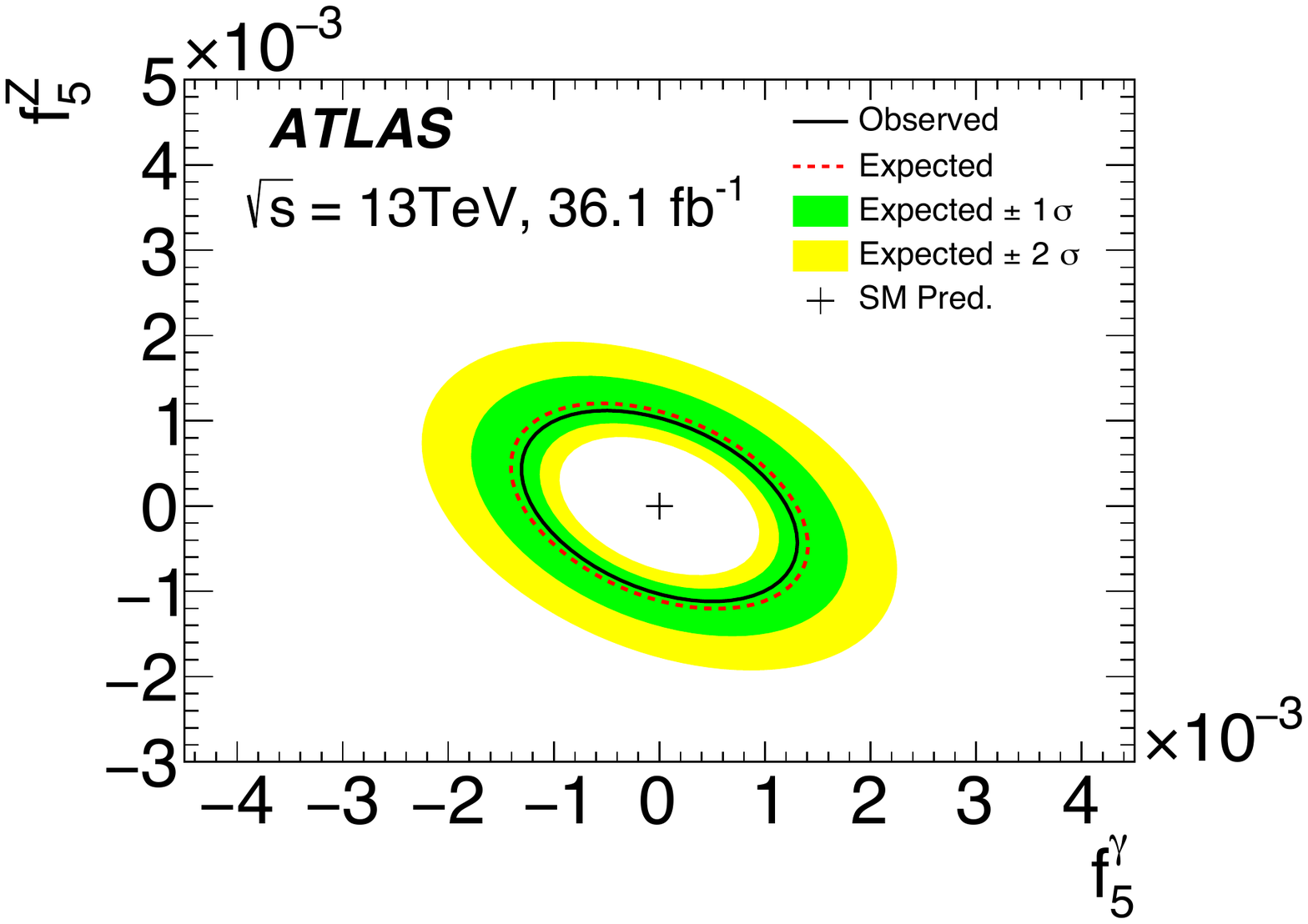}}
\hfill \hfill
\subfloat[]{\raisebox{0.25em}{\includegraphics[height=0.23\textheight]{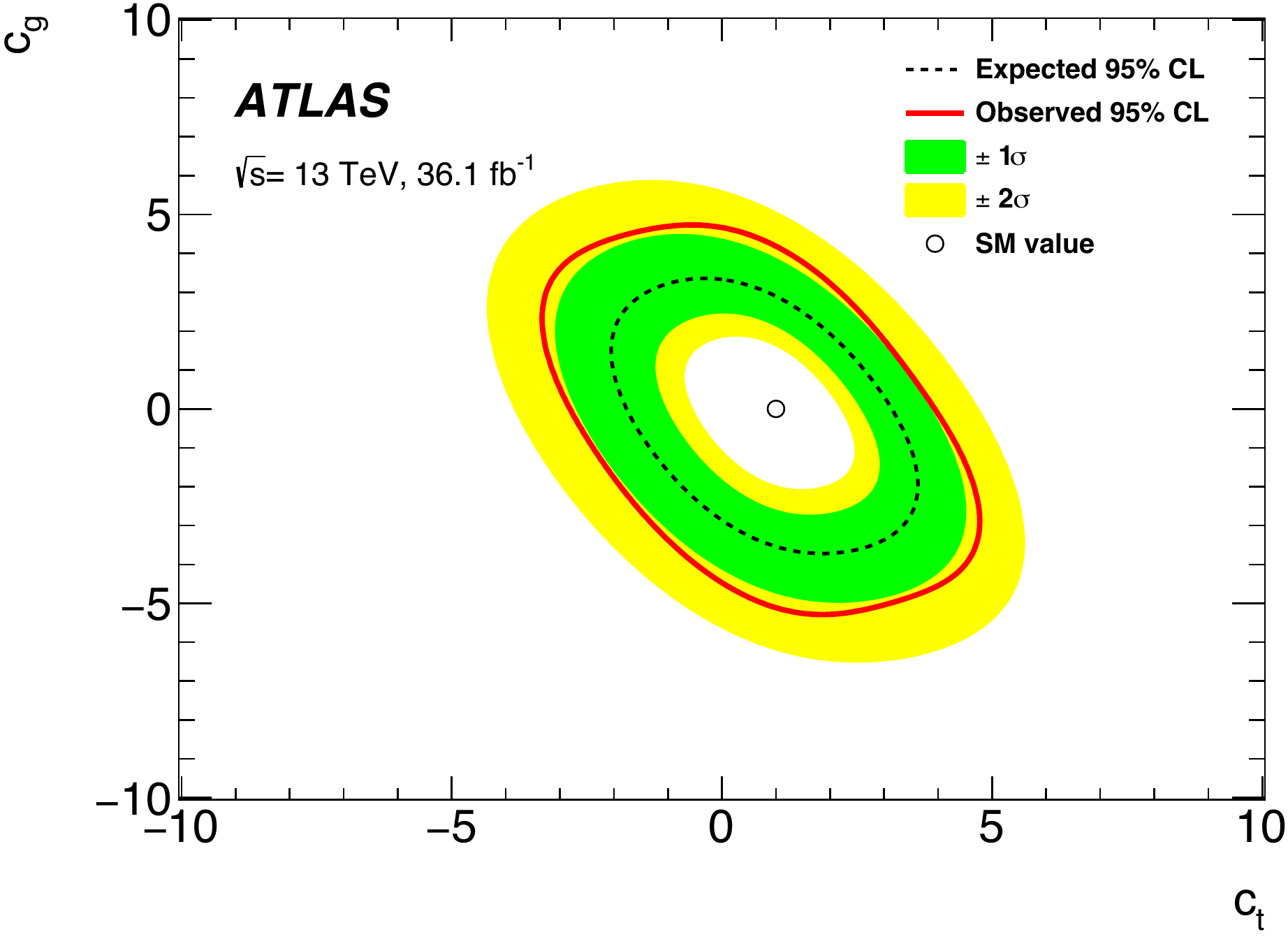}}}
\hfill
\caption{ Observed and expected exclusion limits: 
(a) As function of two $f_5$ aTGC parameters for the $ZZ$ analysis, while the
$f_4$ parameters are set to zero~\cite{ZZinv}.  (b) In the $c_{g}$ versus $c_{t}$ plane for
modified $t\bar{t}H$ and $ggH$ couplings, as obtained in the four-lepton
analysis~\cite{FourLep}.}
\label{fig:EFT}
\end{figure}

Since no significant
deviations from the SM are observed in the $ZZ$ analysis, upper limits are placed on the aTGC
parameters, which typically manifest themselves as a signal
excess growing rapidly as the partonic centre-of-mass energy $\!\!\shat$
increases. In this analysis, aTGCs are searched for using the \ptll spectrum
in the fiducial region, motivated by the fact that \ptll is correlated with
$\!\!\shat$ and has a good experimental resolution.
The contribution
due to aTGCs is introduced using an effective vertex function
approach, including two coupling parameters that violate
$CP$ symmetry, $f_4^\gamma$ and $f_4^Z$, as well as two
$CP$-conserving ones, $f_5^\gamma$ and $f_5^Z$.  
Furthermore, the coupling parameters are used to extract information
about the dimension-eight operators of the effective field
theory. Since the sensitivity to possible aTGCs lies
in the high-\ptll region, only the bins with $\ptll > 150 \GeV$ are
considered in the search. Figure~\ref{fig:ZZ}b compares the measured
\ptll spectrum in this region of phase space to the SM prediction alone and
to the SM prediction augmented with aTGCs corresponding to different values of
the coupling parameters described above.
Two-dimensional confidence intervals (CIs)
for each pair of coupling parameters are derived with the other two parameters
set to zero. Figure~\ref{fig:EFT}a presents the two-dimensional CI
contours for one of the six possible pairs of aTGC coupling parameters.  

In the four-lepton analysis the
detector-corrected four-lepton mass distribution is used to constrain possible
BSM modifications of the couplings of the Higgs boson to top quarks ($c_{t}$)
and gluons ($c_{g}$), which is predicted to be zero in the SM.  On-shell rates for Higgs
production via gluon--gluon fusion are only sensitive to $|c_{t} + c_{g}|^{2}$,
but measurements at higher mass can be used to probe these
parameters independently, as the partonic centre-of-mass energy of the process
becomes larger than the top-quark mass.  This provides an interesting test of
the off-shell behaviour beyond dedicated measurements based on the rare
$t\bar{t}H$ production mode.  
The yield from $\ggZZ$ is parameterised as a function of $c_{t}$ and $c_{g}$.
The observed and expected 
exclusion contours are shown in
Figure~\ref{fig:EFT}b, where the expected limit has green and yellow bands
indicating uncertainties of $1\,\sigma$ and $2\,\sigma$.  The parameter
space which lies outside of the observed contour is excluded. 

\FloatBarrier

\section{Measurements of three massive vector bosons using 2015--2017 data}
\subsection{Evidence for $VVV$ production}

The joint production of three vector bosons is a rare process in the Standard
Model.  Studies of triboson production can test the non-Abelian gauge structure
of the SM theory and any deviations from the SM prediction would provide hints
of new physics at higher energy scales.  Triboson production has been studied
at the LHC using data taken at $\sqrt{s}= 8 \TeV$ for processes such as
$\gamma\gamma\gamma$, \Wgg, $Z\gamma\gamma$, \WWg, \WZg and \WWW.

A search for the production of three massive vector bosons is performed,
extracting one common signal modifier from a simultaneous fit to four selection
regions targeting different channels.  Events with two same-sign leptons and at
least two reconstructed jets are selected to search for $WWW \to \ell \nu \ell
\nu qq$. Events with three leptons without any SFOS lepton pairs are used to
search for $WWW \to \ell \nu \ell\nu \ell \nu$, while events with three leptons
and at least one SFOS lepton pair and one or more reconstructed jets are used
to search for $WWZ \to \ell \nu qq \ell \ell$.  Finally, events with four
leptons are analysed to search for $WWZ \to \ell \nu \ell \nu \ell \ell$ and
$WZZ \to qq \ell \ell \ell \ell$. 

\begin{figure}[htbp]
\centering
\includegraphics[height=0.12\textheight]{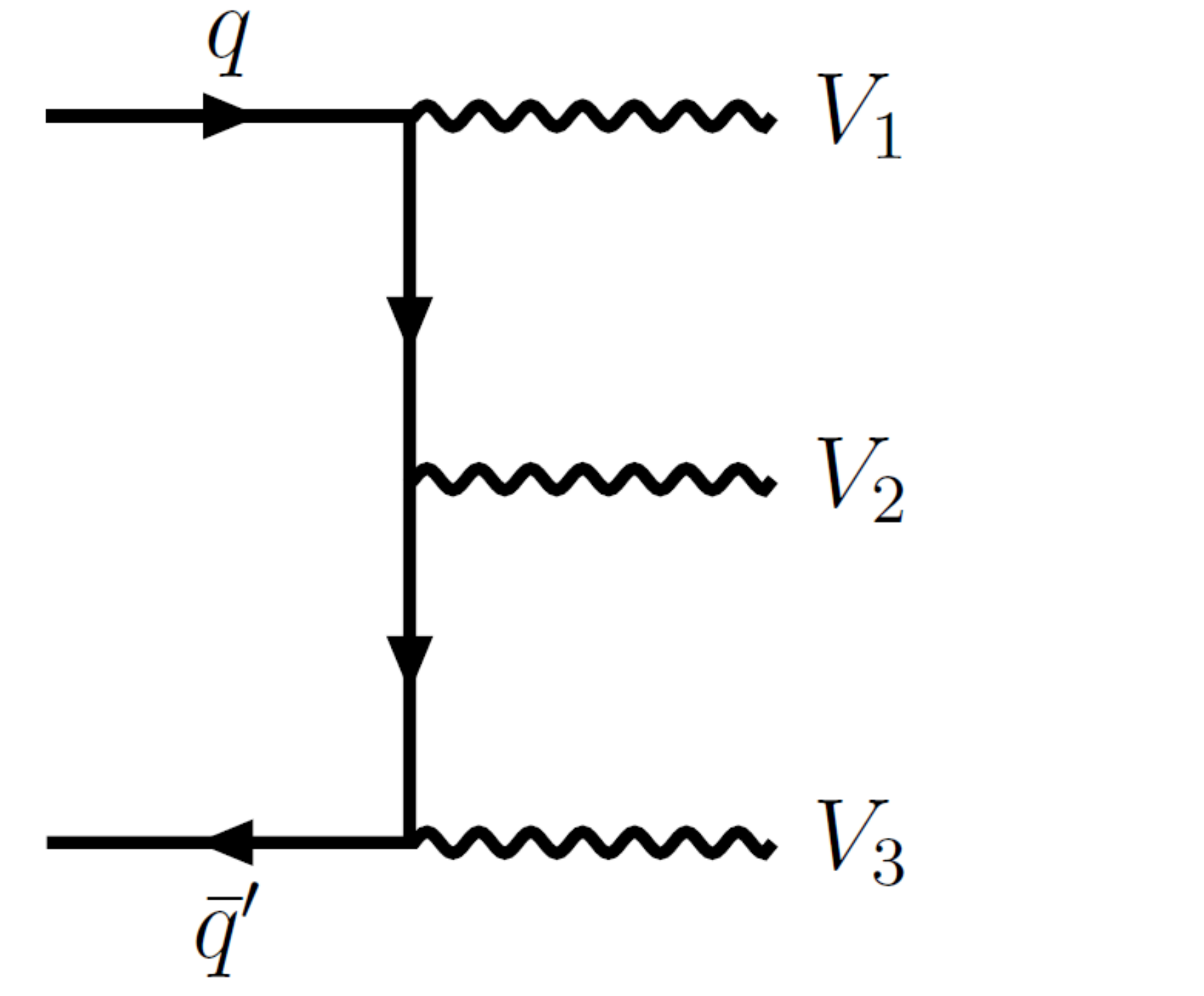}\hfill
\includegraphics[height=0.12\textheight]{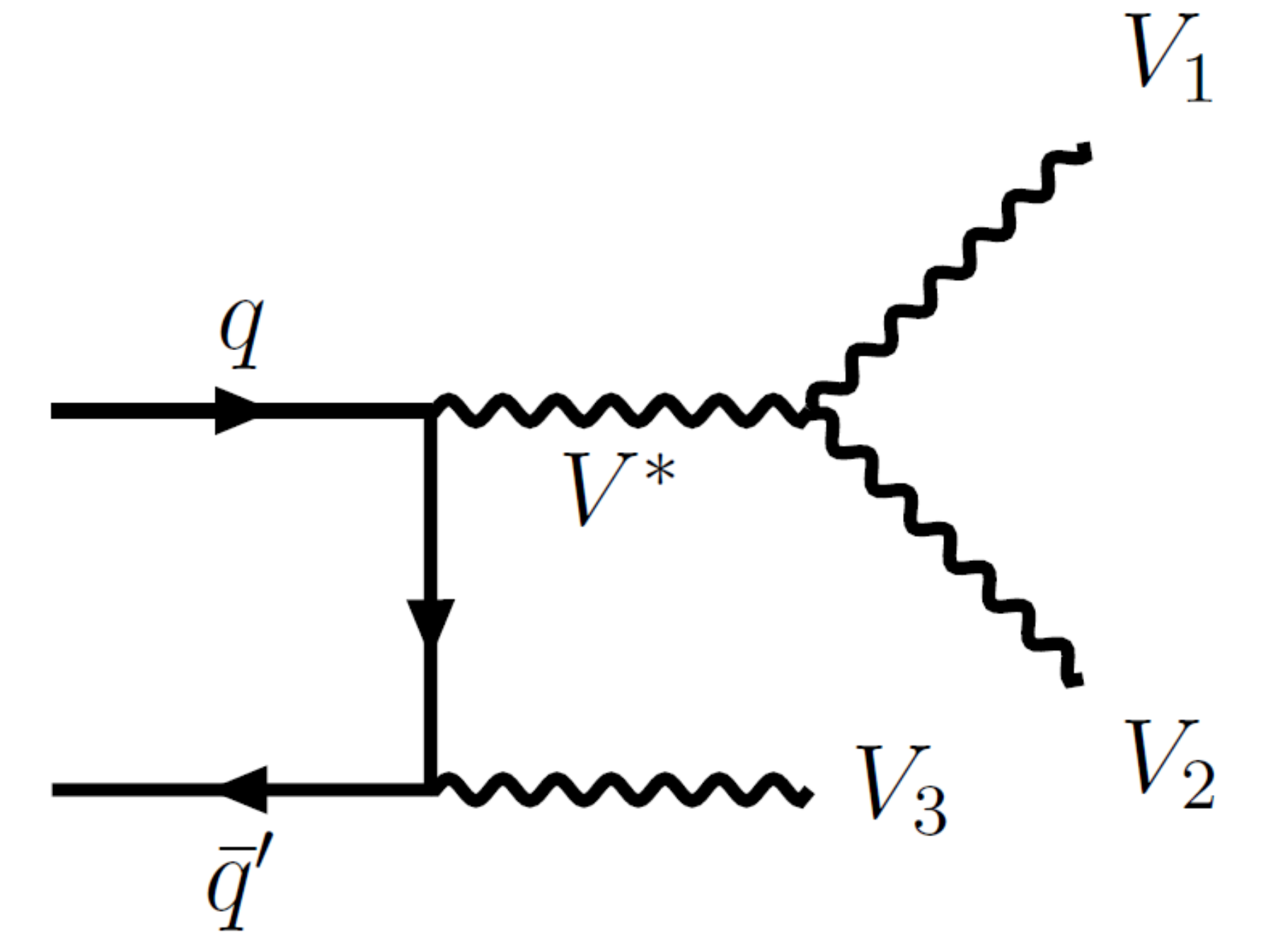}\hfill
\includegraphics[height=0.12\textheight]{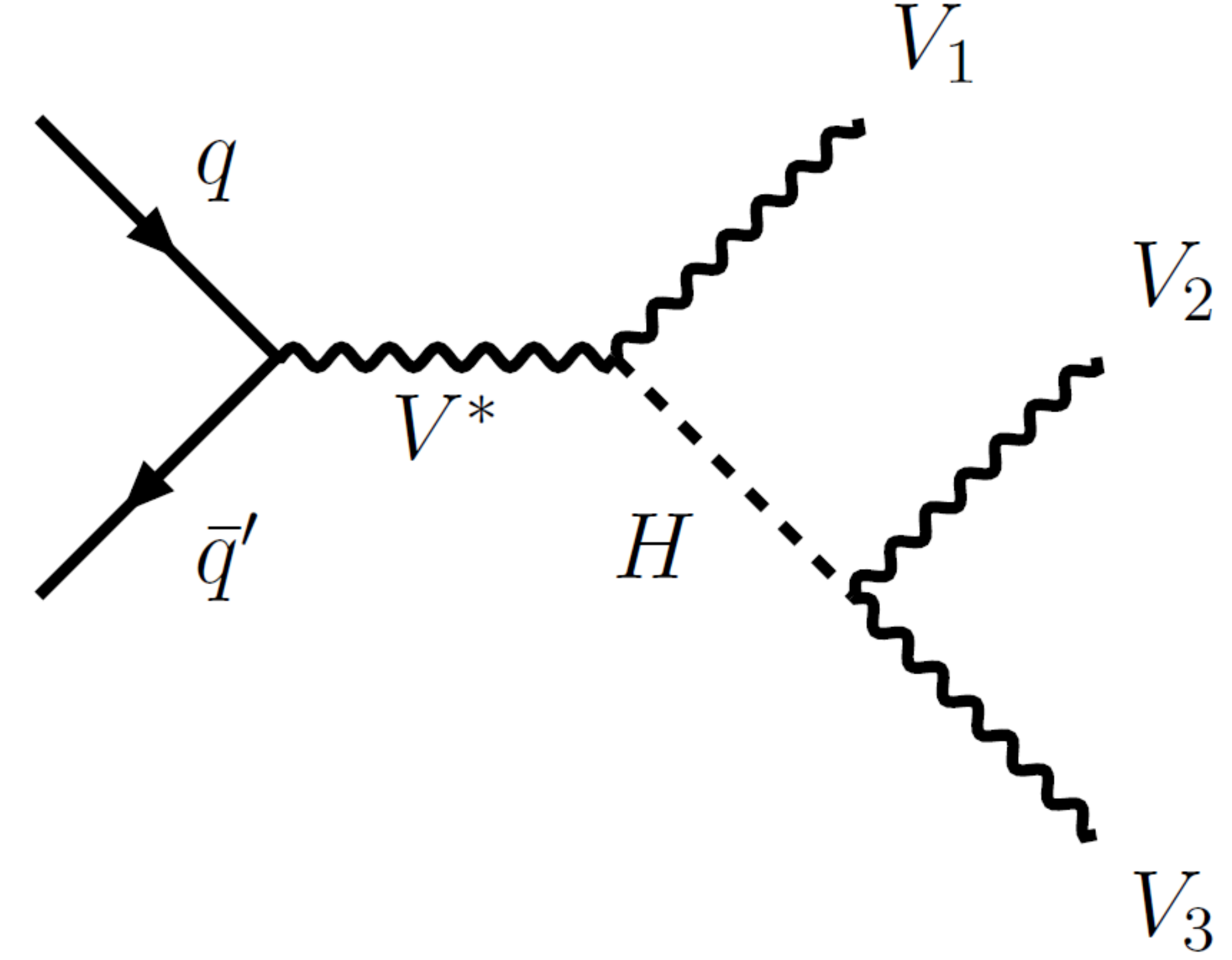}\hfill
\includegraphics[height=0.12\textheight]{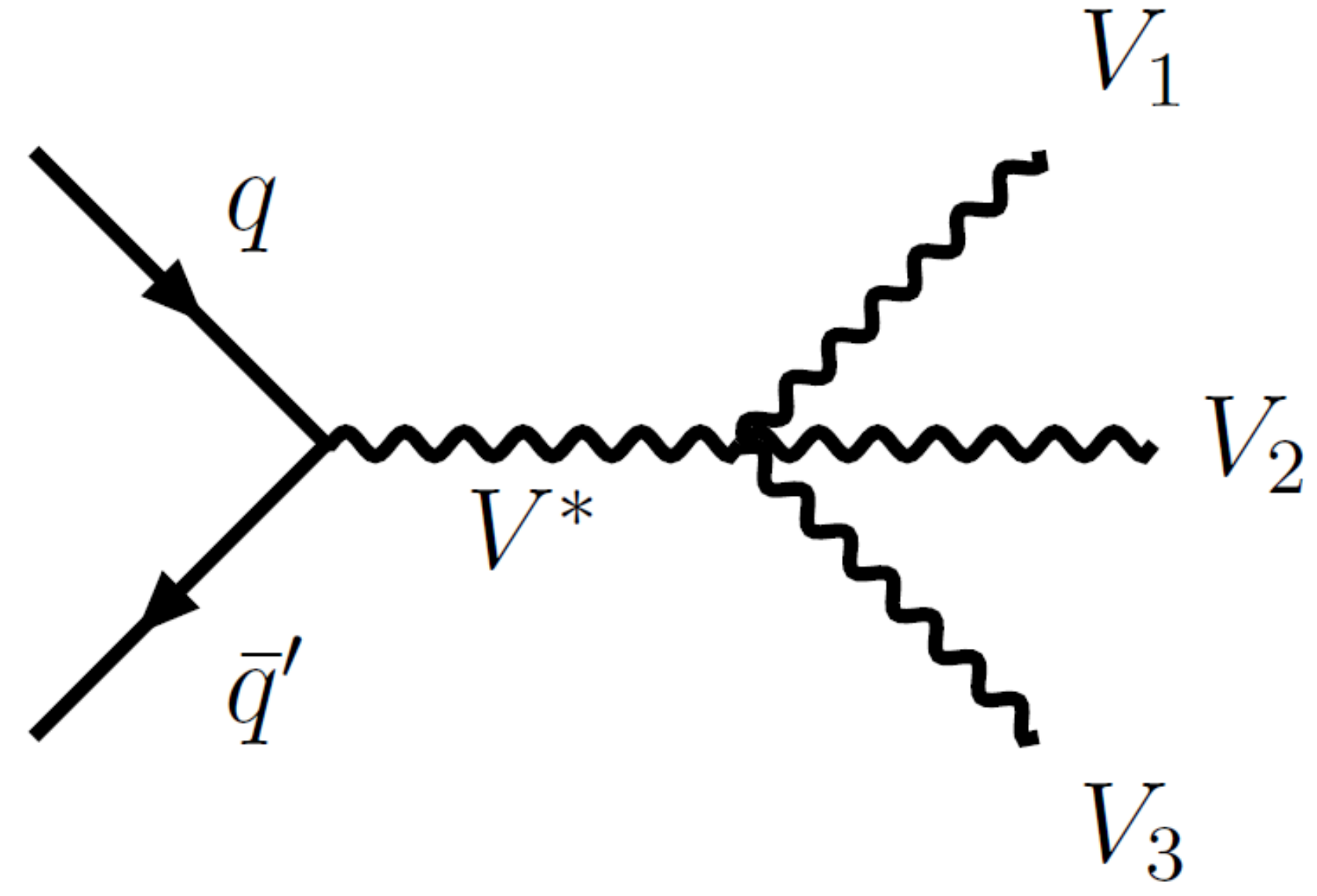}
\caption{Representative Feynman diagrams at LO for the production of three
massive vector bosons, including diagrams sensitive to triple and quartic gauge
couplings.}
\label{fig:WVVfeyn}
\end{figure}

At LO in QCD, the production of three massive vector bosons can proceed via the
radiation of each vector boson from a fermion, from an associated boson
production with an intermediate boson ($W$, $Z/\gamma^*$ or $H$) decaying into
two vector bosons, or from a QGC vertex.  Representative Feynman diagrams are
shown in Figure~\ref{fig:WVVfeyn}.

Triboson signal events were generated using \SHERPA\ where all three bosons are
on-mass-shell. Events with off-mass-shell bosons through $WH \rightarrow WVV^*$
and $ZH \rightarrow ZVV^*$ were generated using \POWHEGBOX or \PYTHIA.  Both
on-mass-shell and off-mass-shell processes were generated at NLO QCD accuracy
and are included in the signal definition. The expected cross sections for
$WWW$ and $WWZ$ production are \SI{0.50}{pb} and \SI{0.29}{pb}, respectively,
with an uncertainty of $\sim\!10\,\%$.

In order to reject leptons likely to be originating from heavy-flavour decays,
``nominal'' leptons have to pass a requirement on a dedicated boosted decision
tree (BDT).  Electrons have to pass an additional BDT requirement to reject
electrons likely to have the electric charge wrongly measured.

To select \ssll candidates, events are required to have exactly two nominal
leptons with the same electric charge, at least two jets, and no identified
$b$-tagged jets. Four regions are considered, based on the lepton flavour,
namely \elel, \elmu, \muel, and \mum.  Requirements on the invariant mass of
the dilepton system are imposed, in order to reduce the contribution from the
$WZ$ process.  Cuts applied on the dijet invariant mass and $\Delta \eta_{jj}$
mainly reduce the contributions from the same-sign $WW$ vector boson scattering
process.  Additionally, in the \elel final state, requirements on \MET and
$m_{\ell\ell}$ are imposed to reduce contamination from $Z \to ee$ where the
charge of one electron is misidentified.  To select \threelep candidates,
events are required to have exactly three nominal leptons and no identified
$b$-tagged jets.  To reduce the contribution from the $WZ$ process, events are
required to have no SFOS lepton pairs, and thus only $\mu^\pm e^\mp e^\mp$ and
$e^\pm \mu^\mp \mu^\mp$ events are selected.

A major background originates from the $WZ$+jets $\to \ell \nu \ell \ell$+jets
process, contributing to the \ssll channel when one lepton is not reconstructed
or identified, or to the \threelep channel, when a $Z$ boson decays into a pair
of $\tau$ leptons both of which decay to an electron or muon.  Simulation
(\SHERPA) is used to estimate this background.  Contributions from SM processes
that produce at least one non-prompt lepton are estimated using a data-driven
method by introducing ``fake'' leptons.  Fake leptons are defined by looser
requirements and are mutually exclusive with the nominal leptons.  Simulation
shows that the \ttbar process is the dominant contributor of events with fake
leptons.  Events containing one (two) nominal lepton(s) and one fake lepton are
scaled by a ``fake factor'' to predict the non-prompt lepton background
contribution in the \ssll (\threelep) channel.  The fake factor is derived from
two \ttbar-enriched regions selected with two or three leptons (no SFOS lepton
pairs) and exactly one $b$-tagged jet.  Events resulting from the $V\gamma jj$
production can pass the signal selection criteria if the photon is
misreconstructed as an electron.  This contribution is evaluated using a
data-driven method similar to the non-prompt lepton background evaluation by
introducing ``photon-like'' electrons.  The charge misidentification background
originates from processes that produce oppositely-charged prompt leptons where
one lepton's charge is misidentified and results in final states with two
same-sign leptons.  The background is estimated using a data-driven technique.

The experimental signature of the \lllwvz, \llllwwz, and \llllwzz processes is
the presence of three or four leptons.  In order to increase the signal
acceptance, a looser lepton definition is also used.  Six regions are defined
with either three or four loose leptons, sensitive to triboson final states
containing $Z$ bosons.  Among all possible SFOS lepton pairs, the one with
$m_{\ell\ell}$ closest to $m_Z$ is defined as the $Z$ candidate.  In all
regions, the presence of such a $Z$ candidate with $|m_{\ell\ell} - m_Z| < 10
\GeV$, is required.  Furthermore, any SFOS lepton pair combination is required
to have a minimum invariant mass of $m_{\ell\ell} > 12 \GeV$.  Events with
$b$-tagged jets are vetoed.  For the three-lepton channel, the lepton which is
not part of the $Z$ candidate is required to be a nominal lepton and the scalar
sum of the transverse momenta of all leptons and jets is required to be large,
significantly reducing the contribution of the $Z\to\ell\ell$ processes with
one additional non-prompt lepton.  Three regions are defined according to the
number of jets in the event: one jet (\TLjone), two jets (\TLjtwo), and at
least three jets (\TLjthree).  For the four-lepton channel, the third and
fourth leading leptons are required to be nominal leptons, and the two leptons
which are not part of the $Z$ candidate definition are required to have
opposite charges. They are used to define three regions, depending on whether
they are different-flavour (\FLDF), or same-flavour and their mass is
compatible with the $Z$ boson mass (\FLSFZ) or not (\FLSFnoZ).

In each of the six regions, the distribution of a dedicated BDT discriminant,
separating the \WVZ signal from the dominating diboson background, is fed as
input to the binned maximum-likelihood fit to extract the signal.  For the
three-lepton channels, 12--15 input variables are used.  The variables are
chosen from a list of discriminating variables, including the trilepton
invariant mass, the invariant mass of different lepton or jet pairs, the
leptons' and jets' \pt, the number of reconstructed jets, the scalar sum of all
leptons' or jets' \pt, \met, \HT and the invariant mass of all leptons, jets
and \met.  For the four-lepton channels, six input variables are used for each
of the final states.  These variables are chosen from the following list: \met,
\HT, the scalar sum of all leptons' \pt, the invariant masses of lepton pairs,
the four-lepton invariant mass, the number of reconstructed jets, and the
scalar sum of all jets' \pT.

Due to the required presence of nominal leptons in the three- and four-lepton
channels, backgrounds with a $Z$ boson and non-prompt leptons are reduced.  The
remaining backgrounds are dominated by processes with prompt leptons and thus
the backgrounds are estimated using simulation.  The \ttZ background is
determined in a region defined like the \TLjthree region with the exception
that no requirement on \HT is applied, and at least four jets are required, of
which at least two are $b$-tagged. This region is included as a single-bin
control region in the fit model.

The \WWW, \WWZ and \WZZ regions are combined using a binned profile likelihood
method based on a simultaneous fit to distributions in the eleven signal
regions (\elel, \elmu, \muel, \mum, \threelep, \TLjone, \TLjtwo, \TLjthree,
\FLDF, \FLSFZ, and \FLSFnoZ) and the \ttZ background control region.  The
distributions used in the fit are the $m_{jj}$ distributions for the \ssll
channel and the full BDT distributions for the \WVZ three-lepton and
four-lepton channels. The number of selected events in the \threelep channel
and the \ttZ control region are each included as a single bin in the fit. In
total, 186 bins are used in the combined fit.  The likelihood function depends
on the signal-strength parameter $\mu$, a multiplicative factor that scales the
number of expected signal events, and a set of nuisance parameters that encode
the effect of systematic uncertainties of the signal and background
expectations, implemented as Gaussian, log-normal or Poisson constraints.  The
same value for $\mu$ is assumed for the on- and off-mass-shell \WWW, \WWZ and
\WZZ processes.  

\begin{figure}[htbp]
\centering
\subfloat[]{\includegraphics[height=0.212\textheight]{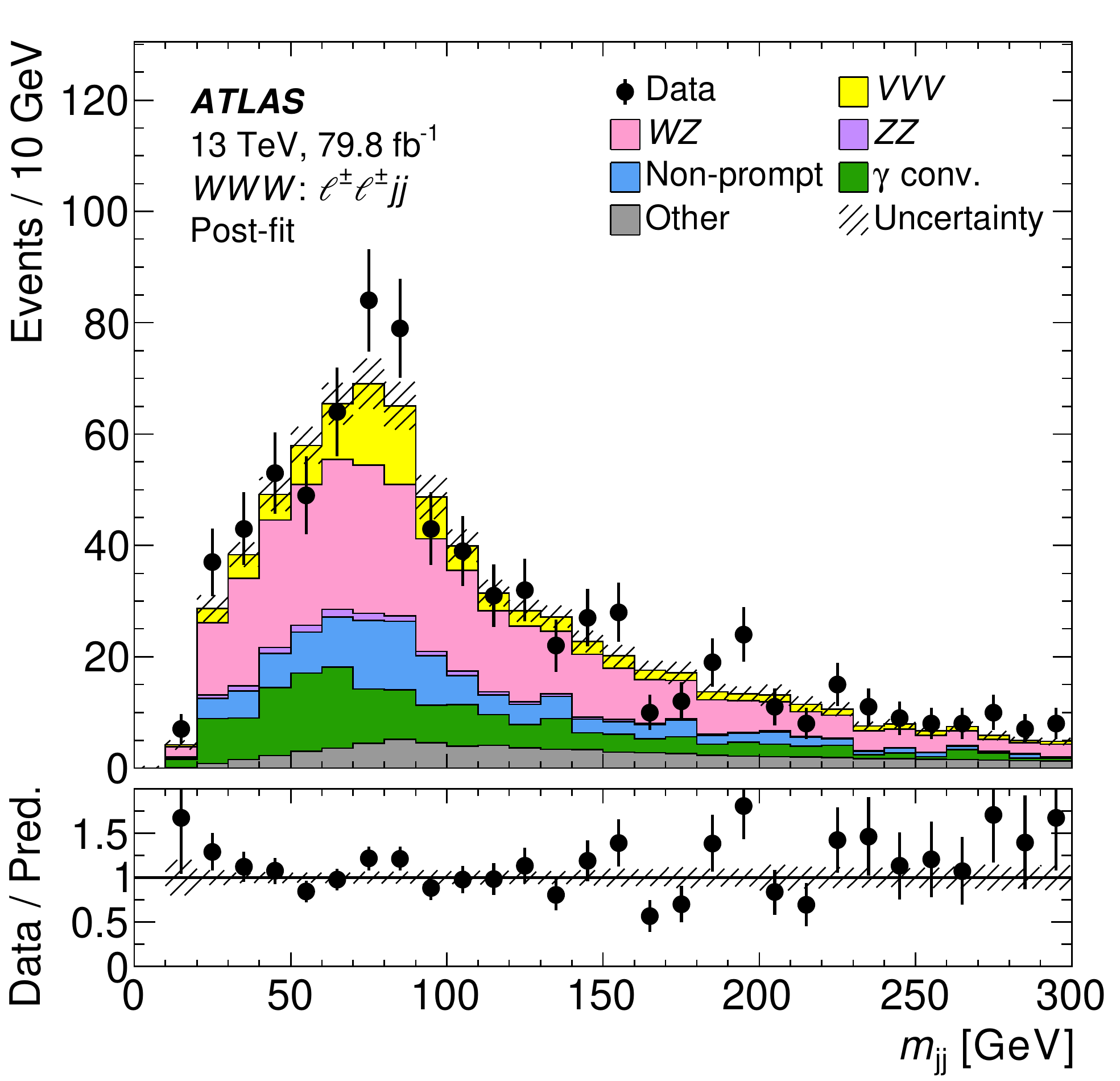}}\hfill
\subfloat[]{\includegraphics[height=0.212\textheight]{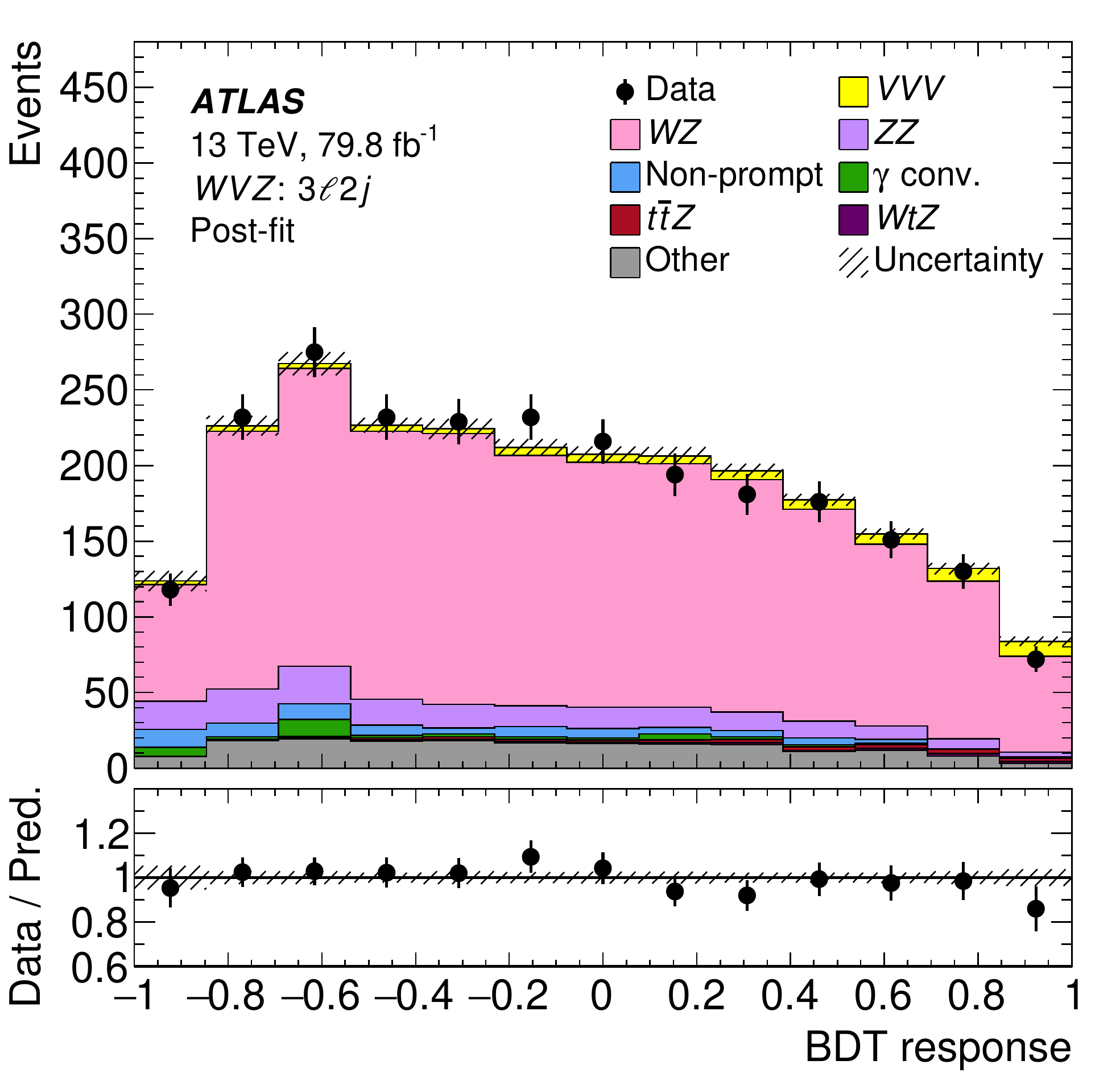}}\hfill
\subfloat[]{\includegraphics[height=0.212\textheight]{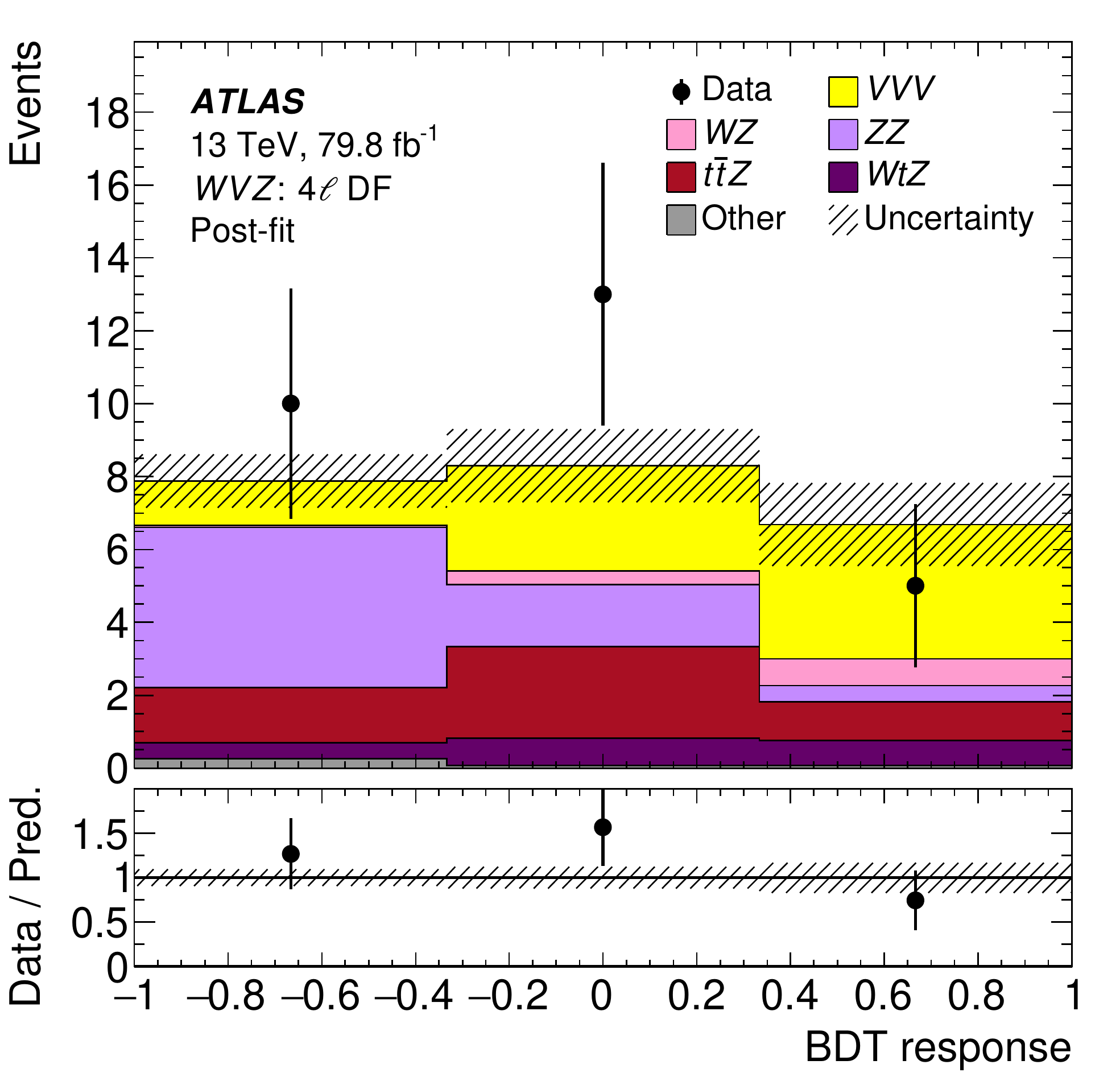}}
\caption{Post-fit distributions~\cite{VVV} of (a) $m_{jj}$ for the \sswww analysis (\elel,
\elmu, \muel, \mum combined), and of the BDT output for the (b) \TLjtwo and (c)
\FLDF channels of the $WVZ$ analysis.}
\label{fig:WVVsign}
\end{figure}

Figure~\ref{fig:WVVsign} shows the comparison between data and post-fit
prediction of the combined $m_{jj}$ distribution for the \lljj channel, and the
BDT output distributions in the \TLjtwo and \FLDF regions for the \WVZ
analysis.  The \TLjtwo and \FLDF regions are chosen since they have the best
sensitivity among the three-lepton and four-lepton channels.  Data and
predictions agree in all distributions.

\begin{figure}[htbp]
\centering
\hfill
\subfloat[]{\raisebox{1em}{\includegraphics[height=0.27\textheight]{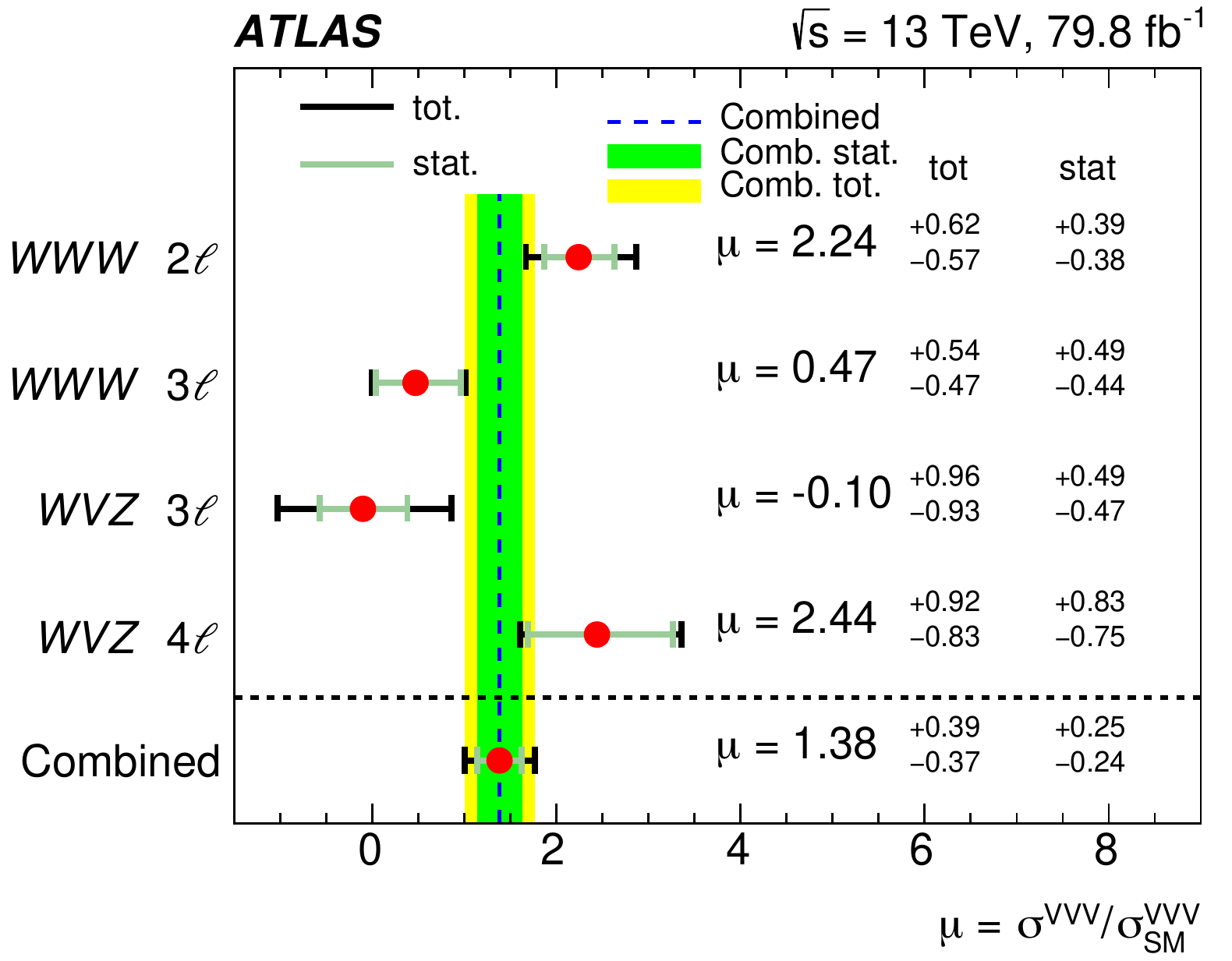}}}
\hfill\hfill
\subfloat[]{\includegraphics[height=0.30\textheight]{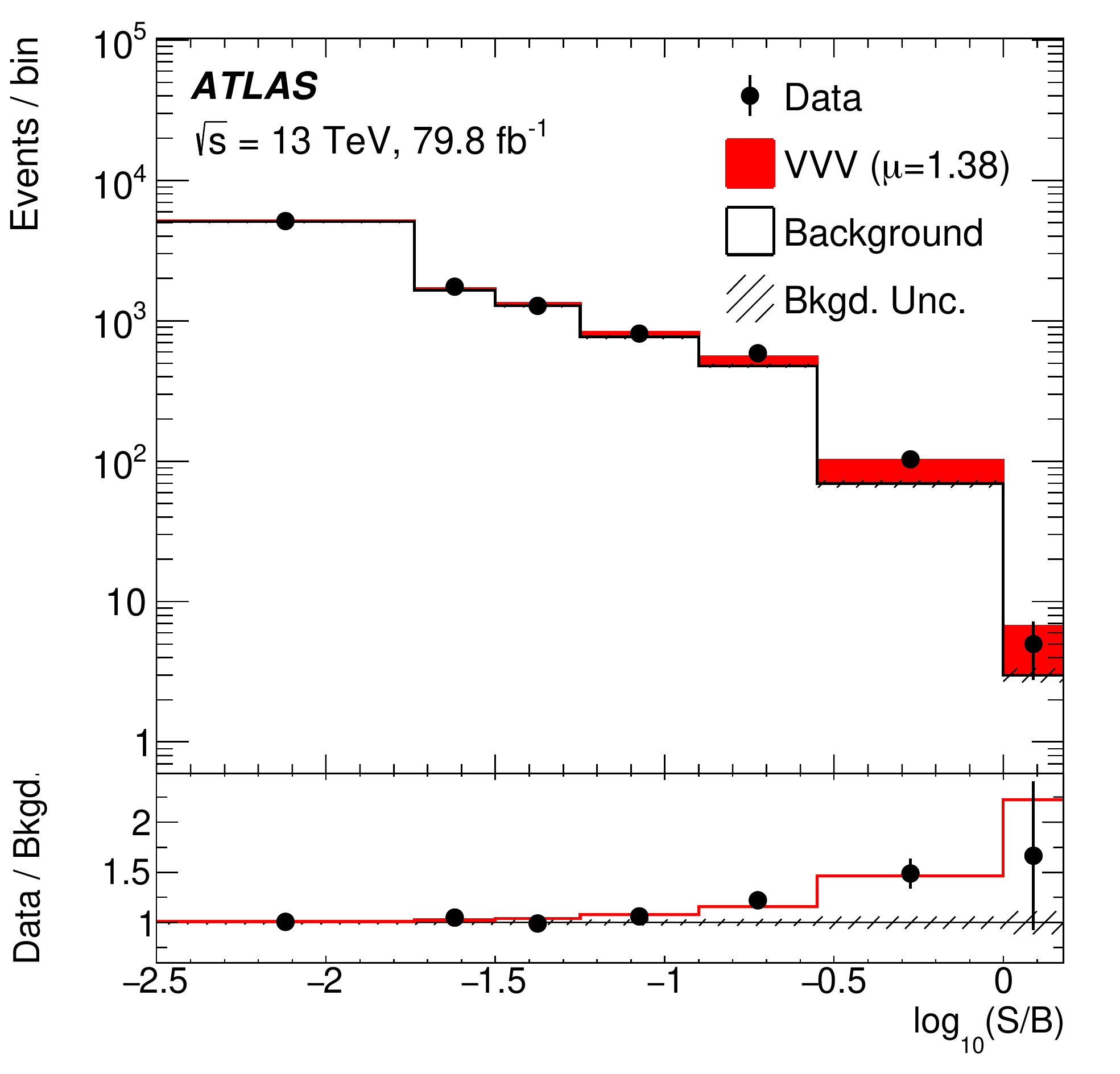}}
\hfill
\caption{(a) Observed values of the signal strength per $VVV$ analysis channel
and combination~\cite{VVV}. (b) All signal regions combined into bins of
$\log_{10}(\text{S/B})$~\cite{VVV}.}
\label{fig:WVVres}
\end{figure}

Figure~\ref{fig:WVVres}a shows the observed value of the signal strength.
Results are shown for the \WWW and \WVZ channels separately, fixing the other
signal to its SM expectation, and combined.  The combined best-fit signal
strength for the $VVV$ process, obtained by the fit to the eleven signal
regions and one control region is \obsvvverr with respect to the SM prediction.
The statistical uncertainty is $^{+0.25}_{-0.24}$ and the systematic
uncertainty is $^{+0.30}_{-0.27}$.  The largest systematic uncertainties come
from uncertainties related to data-driven background evaluations affecting the
\WWW channels and from theoretical uncertainties related to renormalisation and
factorisation scale variations, mostly in the diboson background, evaluated
using simulations.  The observed (expected) significance for $WWW$ production
is $3.3\,\sigma$ ($2.4\,\sigma$), and $2.9\,\sigma$ ($2.0\,\sigma$) for $WVZ$
production.  The overall observed (expected) significance for $VVV$ production
is $4.0\,\sigma$ ($3.1\,\sigma$), constituting evidence for the production of
three massive vector bosons.  Figure~\ref{fig:WVVres}b shows the data,
background and signal yields, where the final-discriminant bins in all signal
regions are combined into bins of $\log_{10}(\text{S/B})$, S being the expected
signal yield and B the background yield. The background and signal yields are
shown after the global signal-plus-background fit to the data.

The measured signal strengths from the fits and their uncertainties are
converted to inclusive cross-section measurements using the signal samples and
the central values of the theoretical predictions.  All uncertainties
determined in the fit are included in the conversion, except for the
normalisation uncertainty in the signal prediction.  The results are:
$\sigma_{WWW} =
0.68^{+0.16}_{-0.15}\,\text{(stat.)}\,^{+0.16}_{-0.15}\,\text{(syst.)}$ pb and
$\sigma_{WWZ} = 0.49 \pm
0.14\,\text{(stat.)}\,^{+0.14}_{-0.13}\,\text{(syst.)}$ pb, in agreement with
the Standard Model predictions.  For the $\sigma_{WWZ}$ extraction, the $WZZ$
normalisation is fixed to the SM expectation. 

\section{Summary}

Five recent ATLAS measurements on multiboson production in $pp$ collisions at
the LHC have been presented.  For the four improved measurements on massive
diboson final states, $WW$, $WZ$, $ZZ$, and four leptons, using 2015--2016
data, the measured fiducial inclusive cross sections agree with SM calculations
at NNLO in QCD, including EW corrections. The total uncertainties of the
measured cross sections are $4.5\%$ ($WZ$), $6.8\%$ ($ZZ$), $7.1\%$ ($WW$) and
$8.6\%$ for the rare $Z\to4\ell$ decay. Differential distributions are compared
with several calculations and generators and measurements agree in general,
with some few exceptions.  Sensitive distributions are used to constrain aTGCs
in the framework of EFTs, improving on existing limits. In the $WZ$ channel,
the helicity states of the bosons are extracted for the first time at a hadron
collider. Using 2015--2017 data, ATLAS also established first evidence for the
production of three massive vector boson with a combined significance of
$4.0\,\sigma$ and total cross-section measurements for $WWW$ and $WWZ$ with
uncertainties of the order $30-40\%$.

\section*{Acknowledgements}

The author would like to thank (in alphabetical order) Ulla Blumenschein, Vadim
Kostyukhin, Bogdan Malaescu, Joany Manjarr\'es Ramos, Matthias Schott, Andrea
Sciandra, Ismet Siral, Alex Tuna and Junjie Zhu in their various roles for
making the first evidence for three massive vector boson production possible. 

This work was partially funded by the European Research Council under the
European Union's Seventh Framework Programme ERC Consolidator Grant Agreement
n.~617185 (TopCoup) and 
by the German Federal Ministry of Education and Research (BMBF) in FSP-103 under 
grant n.~05H15PDCAA.


\begin{thebibliography}{99}
\bibitem{WW} \coll{ATLAS}, \emph{Measurement of fiducial and differential
$W^+W^-$ production cross-sections at $\sqrt{s}=$13 TeV with the ATLAS
detector}, \subm{\EPJC}, \arxiv{1905.04242}{ex}.

\bibitem{WZ} \coll{ATLAS}, \emph{Measurement of $W^{\pm}Z$ production cross
sections and gauge boson polarisation in $pp$ collisions at $\sqrt{s} = 13$ TeV
with the ATLAS detector}, \accept{\EPJC}, \arxiv{1902.05759}{ex}.

\bibitem{ZZinv} \coll{ATLAS}, \emph{Measurement of $ZZ$ production in the
$\ell\ell\nu\nu$ final state with the ATLAS detector in $pp$ collisions at
$\sqrt{s} = 13$ TeV}, \subm{JHEP}, \arxiv{1905.07163}{ex}.

\bibitem{FourLep} \coll{ATLAS}, \emph{Measurement of the four-lepton invariant
mass spectrum in 13 TeV proton-proton collisions with the ATLAS detector}, \pub{JHEP
04 (2019) 048}{10.1007/JHEP04(2019)048}, \arxiv{1902.05892}{ex}.

\bibitem{VVV} \coll{ATLAS}, \emph{Evidence for the production of three massive
vector bosons with the ATLAS detector}, \subm{\PLB}, \arxiv{1903.10415}{ex}.

\end{thebibliography}
\end{document}